\documentclass[aps,showpacs,preprint,preprintnumbers,amsmath,amssymb,nofootinbib]{revtex4-1}
\usepackage{amsmath,amssymb,graphicx}
\usepackage{subfigure}
\usepackage{multirow}
\usepackage{dcolumn}
\usepackage{color}
\usepackage{bm}
\usepackage{siunitx}
\usepackage{slashed}
\newcommand{\bea}{\begin{eqnarray}}
\newcommand{\eea}{\end{eqnarray}}

\def\missE{\slashed E} 

\begin{document}

\title{Probing triple-Higgs productions via $4b2\gamma$ decay channel at a 100 TeV hadron collider}

\author{Chien-Yi Chen$^{1,2,3}$, Qi-Shu Yan$^{4,5,6}$, Xiaoran Zhao$^{4}$, Zhijie Zhao$^{4,7,}$\footnote{Correspondence Author: zhaozhijie12@mails.ucas.ac.cn}, Yi-Ming Zhong$^{8}$
\\$^{1}$ Department of Physics, Brookhaven National Laboratory, Upton, New York 11973, USA
\\$^{2}$ Department of Physics and Astronomy, University of Victoria, Victoria, British Columbia V8P 5C2, Canada
\\$^{3}$ Perimeter Institute for Theoretical Physics, Waterloo, Ontario N2J 2W9, Canada
\\$^{4}$ School of Physical Sciences, University of Chinese Academy of Sciences, Beijing 100049, P. R. China
\\$^{5}$ Center for High-Energy Physics, Peking University, Beijing 100871, P. R. China
\\$^{6}$ Center for future high energy physics, Chinese Academy of Sciences 100049, P. R. China
\\$^{7}$ Department of Physics, University of Siegen, 57068 Siegen, Germany
\\$^{8}$ C.N.~Yang Institute for Theoretical Physics, Stony Brook University, Stony Brook, New York 11794, USA
}

\begin{abstract}
The quartic self-coupling of the Standard Model Higgs boson can only be measured by observing the triple-Higgs production process, but it is challenging for the Large Hadron Collider (LHC) Run 2 or International Linear Collider (ILC) at a few TeV because of its extremely small production rate. In this paper, we present a detailed Monte Carlo simulation study of the triple-Higgs production through gluon fusion at a 100 TeV hadron collider and explore the feasibility of observing this production mode. We focus on the decay channel $HHH\rightarrow b\bar{b}b\bar{b}\gamma\gamma$, investigating detector effects and optimizing the kinematic cuts to discriminate the signal from the backgrounds. Our study shows that, in order to observe the Standard Model triple-Higgs signal, the integrated luminosity of a 100 TeV hadron collider should be greater than $1.8\times 10^4$ ab$^{-1}$. We also explore the dependence of the cross section upon the trilinear ($\lambda_3$) and quartic ($\lambda_4$) self-couplings of the Higgs. We find that, through a search in the triple-Higgs production, the parameters $\lambda_3$  and $\lambda_4$ can be restricted to the ranges $[-1, 5]$ and $[-20, 30]$, respectively. We also examine how new physics can change the production rate of triple-Higgs events. For example, in the singlet extension of the Standard Model, we find that the triple-Higgs production rate can  be increased by a factor of $\mathcal{O}(10)$.
\end{abstract}

\maketitle
\preprint{YITP-SB-15-41}

\section{Introduction}
The discovery of the Higgs boson with a mass of around 125--126 GeV\footnote{We use $m_H=126$ GeV in this study. Recent results from the LHC collaborations suggest $m_H=125$ GeV. This change in $m_H$ barely affects our results.} at the LHC \cite{Aad:2012tfa,Chatrchyan:2012ufa} makes it possible to understand electroweak symmetry breaking (EWSB) in detail. To obtain the full knowledge of EWSB, an important task is to measure the Higgs couplings so as to determine whether its properties agree with the Standard Model (SM) predictions. In particular, the measurement of Higgs self-couplings is crucial because it is the only way to reconstruct and verify the scalar potential \cite{Dawson:2013bba}, which can be directly related to our understanding of baryogenesis \cite{Trodden:1998ym} and vacuum stability. In the second part of this paper, we use the singlet extension of the SM to demonstrate how 
the scalar potential can be affected by new physics. 

In the language of an effective field theory, we can parametrize the Higgs self-interaction Lagrangian as
\begin{eqnarray}
   L\supset -\frac{1}{2}m^2_HH^2-\lambda_3\lambda_{SM}vH^3-\frac{1}{4}\lambda_4\lambda_{SM}H^4 + \cdots , \label{eq1.1}
\end{eqnarray}
where higher-dimensional operators denoted by an ellipsis, like operators $H \partial H \cdot \partial H$ studied in Ref. \cite{He:2015spf} and $H^5$,  are neglected here.  In Eq. (\ref{eq1.1}), $v= 246$ GeV is the Higgs field vacuum expectation value (vev), and $m_H=126$ GeV is the Higgs boson mass. In this Lagrangian, we define two free parameters, $\lambda_3$ and $\lambda_4$, to describe the triple- and quartic-Higgs vertices, respectively:
\begin{eqnarray}
  g_{HHH}=6\lambda_3\lambda_{SM} v,\,\, g_{HHHH}=6\lambda_4\lambda_{SM}.
\end{eqnarray}
In the SM, these two free parameters are equal to $1$, i.e., $\lambda_3=\lambda_4=1$, and all higher-dimensional operators vanish. The self-coupling parameter $\lambda_{SM}$ is related to $m_H$ by $\lambda_{SM}=m^2_H/2v^2$. Because of the fact that $\lambda_{SM}\approx 0.13$, the range of $\lambda_4$ can be taken to be around $20$ (its sign is undetermined) in order to guarantee either the validity of the perturbation method or the unitary bound.

Recently, the di-Higgs production at LHC \cite{Plehn:1996wb,Baur:2002qd,Baglio:2012np,Li:2013flc,Bhattacherjee:2014bca} has been a hot topic due to its sensitivity to $g_{HHH}$ and $\lambda_3$. It is well-known that gluon fusion is the dominant process for di-Higgs production at the LHC, and decay channels like $b\bar{b}\gamma\gamma$ \cite{Baur:2003gp,Yao:2013ika}, $b\bar{b}\tau\tau$ \cite{Dolan:2012rv,Barr:2013tda}, $b\bar{b}WW$ \cite{Papaefstathiou:2012qe}, and $b\bar{b}b\bar{b}$ \cite{deLima:2014dta} have been well studied. Previous studies show that the triple self-coupling can be measured within $40\%$ accuracy at LHC Run 2~\cite{Baglio:2012np,Barger:2013jfa}. The double- Higgs production at a 100 TeV hadron collider has also been studied \cite{Barr:2014sga,Papaefstathiou:2015iba}. A study on $HH\rightarrow WW^*WW^*$ shows that the sensitivity can reach up to $13\sigma$ in the SM \cite{Li:2015yia}.

In contrast, very little attention has been paid to triple-Higgs production. Early work on triple-Higgs production has shown that in the SM it is very challenging to discover the signals at $e^+e^-$ colliders, because the cross section of $e^+e^-\rightarrow ZHHH$ is very small. For example, the cross section is only $0.4$ ab at $\sqrt{s}=1$ TeV \cite{Djouadi:1999gv} and  the total production is just 1.2 events for a designed integral luminosity $3$ ab$^{-1}$. However, the triple-Higgs production rate can be enhanced dramatically if there is an extended Higgs sector. The cross section of triple-Higgs production can be at $\mathcal O (0.1)$ pb in the two-Higgs-doublet Model \cite{Ferrera:2007sp,Ferrera:2008nu}. So the triple-Higgs production at $e^+e^-$ colliders is an important process to probe new physics. It is also remarkable that the Higgs self-couplings could be measured to some degree  via indirect or loop processes at $e^+ e^-$ colliders~\cite{McCullough:2013rea}.

The cross section of triple-Higgs production at hadron colliders was calculated in Refs. \cite{Plehn:2005nk,Binoth:2006ym}. Its SM value, via gluon fusion, is  $\mathcal O(0.01)$ fb at the 14 TeV LHC, which is too small to be observed with the current designed luminosity. Moreover, the dominant contribution of this process is the top-loop pentagon diagram \cite{Binoth:2006ym}, which suggests that measurement of $\lambda_4$ is very challenging even if the triple-Higgs production is discovered. [$\lambda_4$ can be read out from the fit cross section given in Eq. (\ref{xsl4}).] In this case, the top mass effect is crucial and leads to a $K$ factor which is similar to the di-Higgs case. A more precise prediction of triple-Higgs production at 100 TeV can be found in Ref. \cite{Maltoni:2014eza}, where it is shown that the cross section can be increased from $3$ to $5$ fb after taking into account the next-to-leading-order (NLO) corrections. 

If we can suppress the SM backgrounds effectively or increase the integrated luminosity enough, it is still possible to observe this process at a 100 TeV machine. Recently, the channel $p p \to HHH\to b\bar{b}b\bar{b}\gamma\gamma$ at the hadron level (with part of  detector simulations implemented) is studied in Ref. \cite{Papaefstathiou:2015paa}. We will comment on it in Sec.~\ref{discussion}.

Although the cross sections of triple-Higgs production have been studied, to our knowledge, serious feasibility studies are still absent in the literature. In this paper, we will focus on the feasibility of  triple-Higgs production at a future 100 TeV hadron collider via  $b\bar{b}b\bar{b}\gamma\gamma$ so as to fill this gap. We include detector simulations by using DELPHES 3.0 \cite{Ovyn:2009tx,deFavereau:2013fsa}. We explore the following three questions related to the physics of a 100 TeV collider:
\begin{enumerate}
\item What is the minimal luminosity to observe the signature of triple-Higgs production via the $4b2\gamma$\footnote{We use the shorthand, for example, $2b$ or $4b$ to denote $b\bar{b}$ or $b\bar{b}b\bar{b}$, respectively.} final state in the Standard Model at a 100 TeV collider after taking into account more realistic detector effects? 
\item What are the bounds on the trilinear and quartic couplings $\lambda_3$ and $\lambda_4$ defined in Eq. (\ref{eq1.1}) that we can achieve by using the triple Higgs production signature?
\item What is the potential to discover new physics via the observation of the final states of triple Higgs bosons? We will use the singlet+SM model as an example to demonstrate this potential.
\end{enumerate}


The structure of this paper is organized as follows. In Sec. \uppercase\expandafter{\romannumeral2}, we describe our Monte Carlo (MC) simulation method. Our analysis is mainly demonstrated in Sec. \uppercase\expandafter{\romannumeral3}. The SM results are presented as a standard candle, and the kinematic cuts are explored and exposed. We also apply two multivariate analysis methods to improve the signal and background discrimination. Based on those analysis methods, we can determine the integrated luminosity for discovering the triple-Higgs boson final states. In Sec. \uppercase\expandafter{\romannumeral4}, the sensitivity of Higgs quartic couplings in the effective Lagrangian are addressed. In Sec. \uppercase\expandafter{\romannumeral5}, the triple-Higgs production in the singlet+SM model is presented. We end this work with some discussions and future outlook.

\section{MC simulation}
 
We use MadLoop/aMC@NLO \cite{Pittau:2012fn} and GoSam \cite{Cullen:2014yla} to generate the matrix elements of triple-Higgs production via gluon fusion. Then we use the VBFNLO code \cite{Arnold:2008rz,Arnold:2011wj,Baglio:2014uba} to perform the phase-space integration, where we set the parton distribution functions as CTEQ6L1 \cite{Pumplin:2002vw}. 

As a cross check, our code yields a cross section $\sigma_\text{14 TeV}=6.67\times 10^{-2}$ fb for the same parameters given by Ref. \cite{Binoth:2006ym}. The two results agree. To arrive at this result, we choose the phase space cuts for the final Higgs bosons as $|\eta(H)|<5.0$ and $P_t(H)>1$ GeV. Then, we set both the renormalization scale and the factorization scale to be the invariant mass of the final states. Our code also performs a reweighting in order to generate unweighted parton-level events. After finishing these cross checks, we use our code to generate unweighted parton-level signal events at the center-of-mass energy of 100 TeV. We use the DECAY package provided by MadGraph 5 to decay Higgs into $b\bar{b}b\bar{b}\gamma\gamma$ final state. Then, we pass each event to PYTHIA 6.4 \cite{Sjostrand:2006za} to simulate the parton shower and to perform hadronization and further decays. 

The parton-level background events are generated by MadGraph5/aMC@NLO \cite{Alwall:2014hca} directly and showered through PYTHIA 8 \cite{Sjostrand:2014zea}. In this paper, we only consider events with at least two tagged $b$ jets, i.e., the $n_b\geq2$ case (cases with a different number of tagged $b$ jets are discussed in Sec. \uppercase\expandafter{\romannumeral6}). Then we take into account two types of dominant background events: $pp\rightarrow b\bar{b}jj\gamma\gamma$ and $pp\rightarrow H t\bar{t}$. To generate the most relevant events, several generator-level cuts are applied for $pp\rightarrow b\bar{b}jj\gamma\gamma$ event generation: for $b$ jets, $P_t(b)>30$ GeV and $|\eta(b)|<5.0$; for other jets, $P_t(j)>20$ GeV, $|\eta(j)|<5.0$; and for $\gamma$'s,$P_t(\gamma)>30$ GeV, $|\eta(\gamma)|<2.5$ and  $|M_{\gamma\gamma}-126$ GeV$|<15$ GeV, where $M_{\gamma\gamma}$ is the invariant mass of two photons. After those cuts, the cross section of $pp\rightarrow b\bar{b}jj\gamma\gamma$,  $\sigma_{b1}$, is 192.8 fb. We do not introduce any extra generator level cuts for the Higgs or tops in the event generation of $pp\rightarrow Ht\bar{t}$. We also require a resonant decay from Higgs to $\gamma\gamma$ when the events are passed to PYTHIA 8.  The cross section of $pp\rightarrow H(\gamma\gamma)t\bar{t}$, $\sigma_{b2}$, with a branching ratio $BR(H\rightarrow \gamma\gamma)\approx 0.25\%$, is found to be 68.2 fb. 

To reduce the fluctuation effects from the MC simulation, we generate 50,000, 150,000, and 150,000 events for the signal, $pp\rightarrow b\bar{b}jj\gamma\gamma$ background, and $H(\gamma\gamma)t\bar{t}$ background, respectively. 

We use FASTJET \cite{Cacciari:2011ma} for jet clustering.  Jets are clustered by using the anti-$k_t$ algorithm \cite{Cacciari:2008gp} with a cone of radius $R=0.5$ and minimum $P_t (j)=30$ GeV. For photon identification, the maximum of isolation efficiency is $95\%$, with transverse momentum $P_t(\gamma)>10$ GeV and $|\eta(\gamma)|\leq 2.5$. The efficiency decreases to $85\%$ for $2.5<|\eta(\gamma)|\leq 5.0$. Pileup effects are neglected in this work. The detector simulation is performed by DELPHES 3.0 \cite{Ovyn:2009tx,deFavereau:2013fsa}. Details about the setup are shown in Appendix~\ref{detectorsimulation}.

The $b$ tagging is simulated by assuming a $60\%$ $b$-jet efficiency working point. The (mis)tagging efficiencies vary with respect to different $P_t$ and $\eta$ of jets. The efficiency curves are given in Appendix~\ref{btag}.
For $P_t(j) =120$ GeV, the $b$-tagging efficiencies for ($b$, $c$, light) jets are (0.6, 0.1, 0.001). Those efficiencies dramatically drop down to (0.28, 0.046, 0.001) at $P_t(j)=30$ GeV. 

We neglect the background events from the processes $pp\rightarrow HW^+W^-$, because $W^{\pm}$ is unable to decay to $b$ quarks, and these background events can be efficiently rejected by two $b$ taggings and its production cross section is much smaller than the process $pp\to t \bar{t} H$. We also neglect the process $pp\rightarrow HZZ$. It has a cross section $\sigma_{HZZ}=29.3$ fb, but its branching ratio of $HZZ\rightarrow \gamma\gamma b\bar{b}b\bar{b}$ is smaller than $0.006\%$. The other backgrounds like $Hb\bar{b}b\bar{b}$ and $b\bar{b}b\bar{b}\gamma\gamma$ can be safely neglected for their small cross sections when compared with the process $pp\rightarrow b\bar{b}jj\gamma\gamma$. We also neglect the background process $pp \to H H jj$ because the cross section is much smaller than those of two dominant background processes we considered here. 

\section{Analysis of the SM}
\subsection{Parton-level Distributions}
\begin{figure}[htbp]
  \centering
  \subfigure{
  \label{Fig1.sub.1}\thesubfigure
  \includegraphics[width=0.4\textwidth]{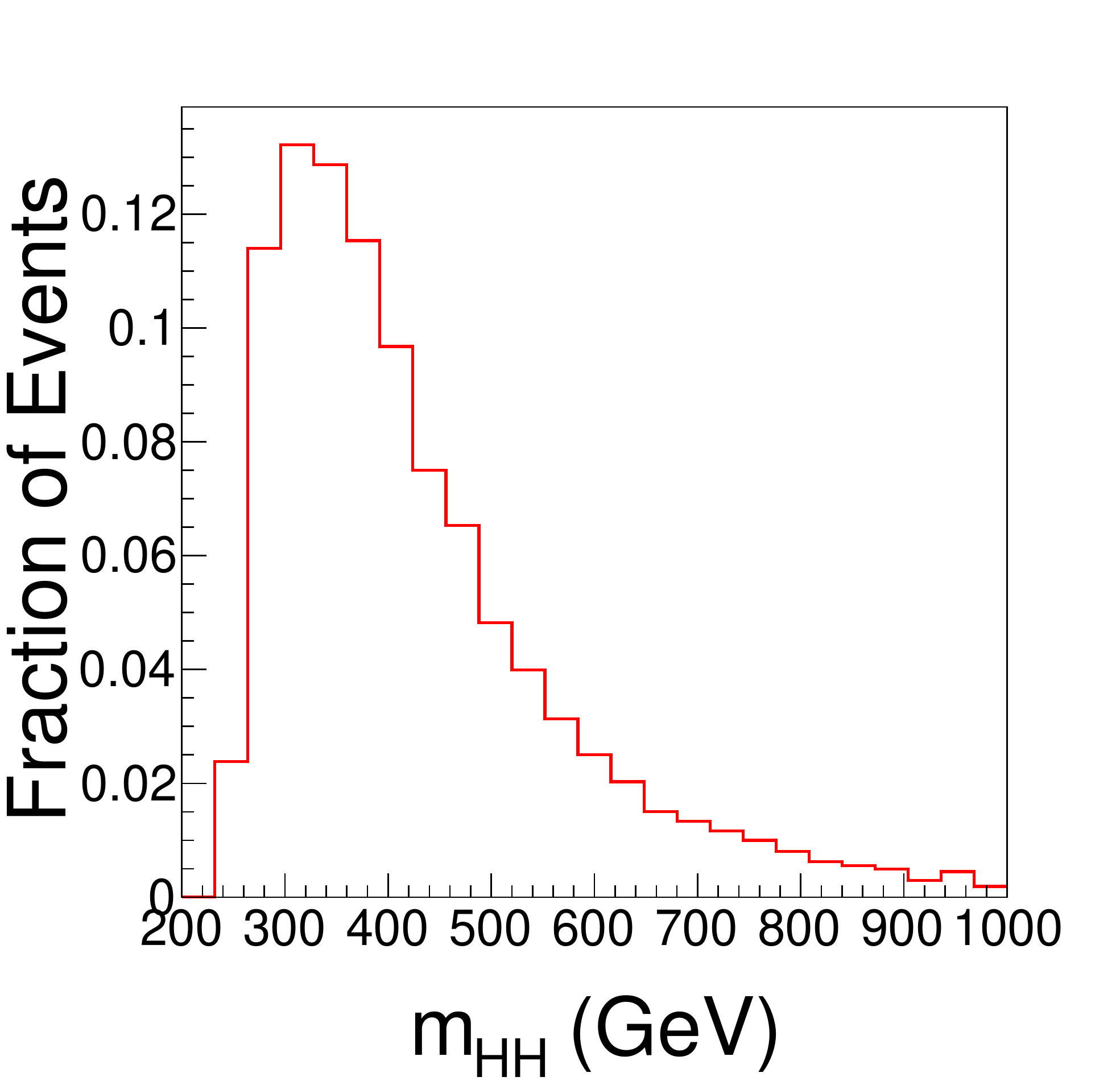}}
  \subfigure{
  \label{Fig1.sub.2}\thesubfigure
  \includegraphics[width=0.4\textwidth]{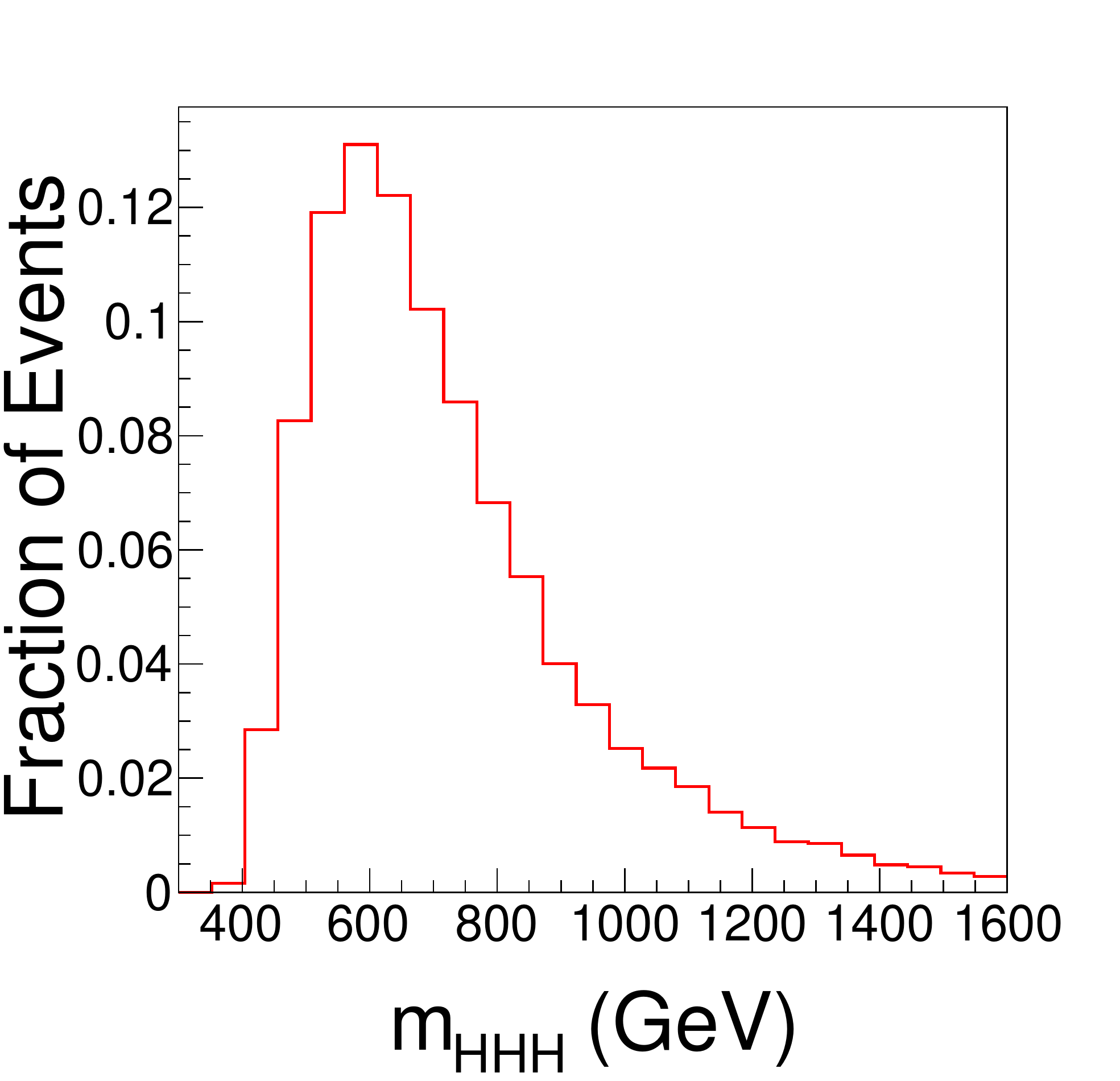}}
  \caption{Distributions of the (a) invariant mass of two Higgs $m_{HH}$ and (b) invariant mass of three Higgs $m_{HHH}$ at the leading-order parton level are shown.}\label{fig1}
\end{figure}

The leading-order cross section of $gg\rightarrow HHH$ in the SM is $\sigma_s=3.05$ fb at a 100 TeV collider. The invariant mass of a pair of Higgs boson $m_{HH}$ in each event and the invariant mass of final states $m_{HHH}$ distributions are shown in Fig. \ref{fig1}. The NLO corrections for this process is large. Therefore, throughout this paper, we assume that the $K$ factor is 2.0 \cite{Papaefstathiou:2015paa}. The peaks of $m_{HH}$ and $m_{HHH}$ are around $350$ and $600$ GeV, respectively. The dominant contributions are from box and pentagon diagrams as we will explain in the next section from our fit by Eq. (\ref{xsl34}). It is noticed that there are long tails in these distributions due to the high center-of-mass energy.
\subsection{Detector-level analysis}
\begin{figure}[htbp]
  \centering
  \subfigure{
  \label{Fig2.sub.1}\thesubfigure
  \includegraphics[width=0.4\textwidth]{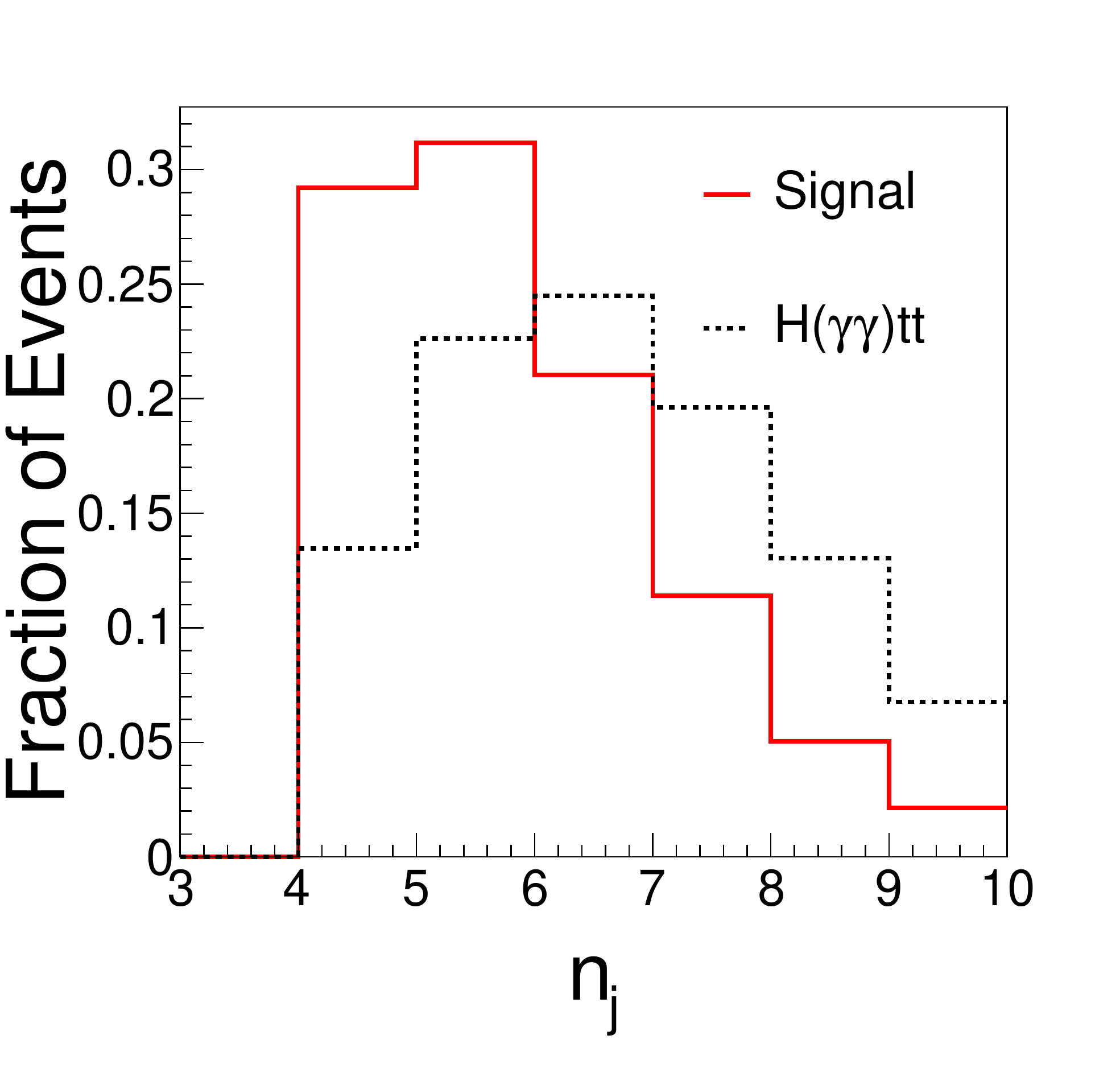}}
  \subfigure{
  \label{Fig2.sub.2}\thesubfigure
  \includegraphics[width=0.4\textwidth]{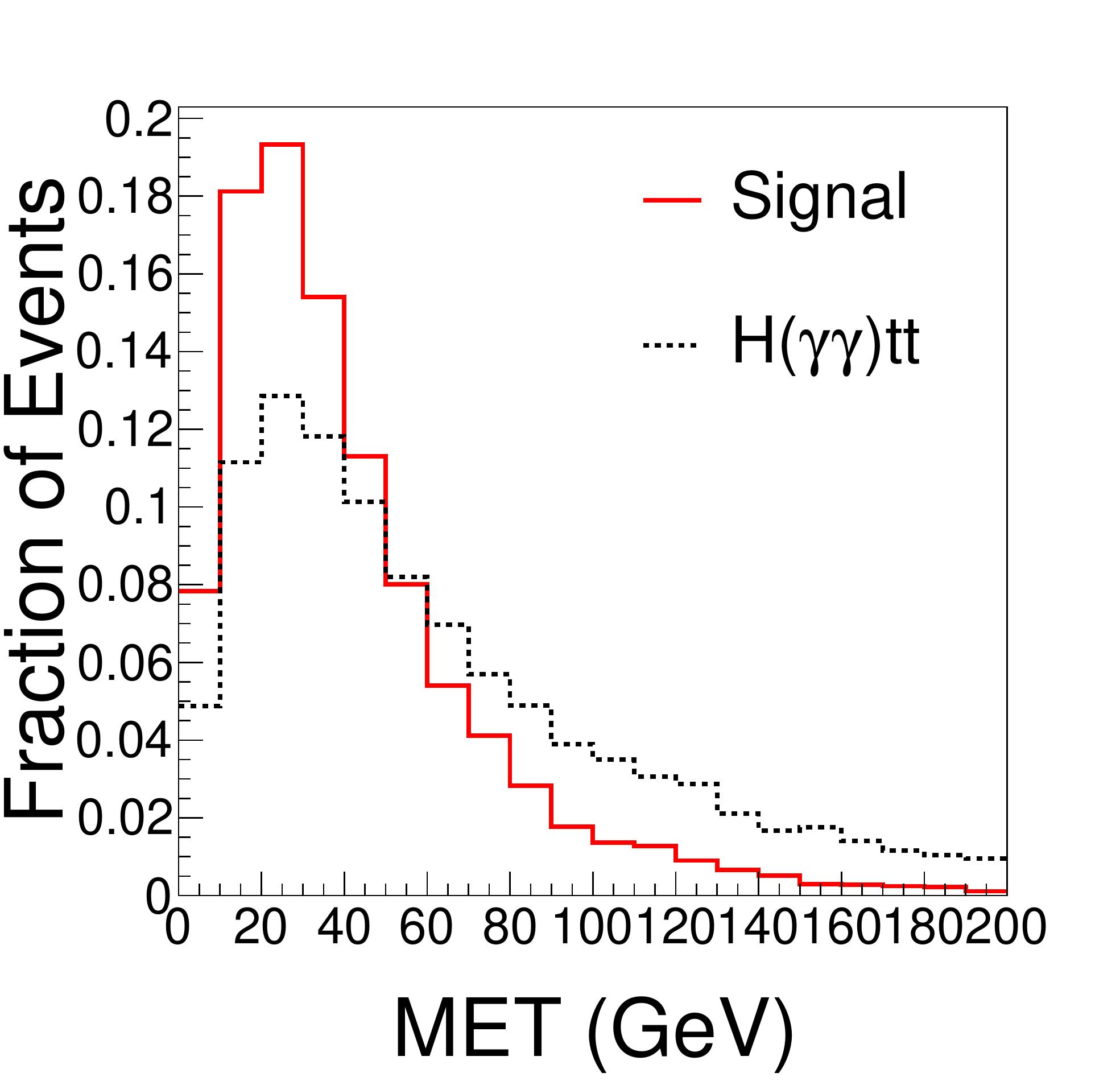}}
  \caption{The distributions of the (a) number of jets for the fully hadronic final states and (b) missing energy transverse for the semileptonic and dileptonic final states for both the signal and the background $p p \to t \bar{t} H $ at the detector level are demonstrated. }\label{fig2}
\end{figure}

Below we focus our analysis on channel $g g\rightarrow H H H\rightarrow b\bar{b}b\bar{b}\gamma\gamma$, which possesses a branching ratio $\approx 0.15\%$ in the all decay final states. To suppress the huge background events and select the most relevant events, we introduce several preselection cuts listed below.
\begin{enumerate}
  \item Only the events with four or five jets are considered, including at least two tagged $b$ jets. The transverse momentum of jets is required, $P_t(j)>30$ GeV. 
  \item The events with exactly two isolated photons with $P_t(\gamma)>30$ GeV are selected. 
\item  For the $ p p \to t \bar{t} H$ background with fully hadronic $t\bar{t}$ decays, where the top quark decays to $b$ and $W^+$, we require that the number of jets reconstructed by the detector should be no more than five. The distribution of the number of jets for this type of background is shown in Fig. \ref{Fig2.sub.1}, which explains why we only consider events with four and five jets.
  \item For the $ p p \to t \bar{t} H$ background with semileptonic and dileptonic $t\bar{t}$ decays, where $W^\pm$ decays to the lepton and neutrino, the detector can reconstruct leptons and a large missing transverse energy (MET). To suppress these two types of backgrounds, we veto the events with any leptons. Details about the detector simulation for leptons are shown in Appendix~\ref{detectorsimulation}. As the leptons and all other visible objects are reconstructed, the MET can be reconstructed. The distribution of MET is shown in Fig. \ref{Fig2.sub.2}, where one can clearly see that the background has a large MET. However, the MET of $Ht\bar{t}$ events are typically much larger than the signal, so the events with MET $>50$ GeV are vetoed.
\end{enumerate}

We would like to make one comment on the first two cuts. These two cuts are quite essential in order to suppress the QCD background from the processes $ p p \to 4 j 2 \gamma$. The cross section of the cross section is computed by the package alpgen \cite{Alpgen}, which yields a result 14.6 pb. After imposing the mass window cut $ 110 \textrm{GeV} < m_{\gamma \gamma} < 140 $ GeV, the cross section of $p p \to 4 j 2 \gamma$ is around 2.3 pb, which is still around ten times larger than the main background $ p p \to 2b 2j 2 \gamma$. But after requiring at least two tagged b jets, this type of background without charm is suppressed by a factor $10^{-5}$, and the total cross section of the background is less than 2 fb, which is less than $2\%$ of the main background $ p p \to 2b 2j 2\gamma$ in our analysis. The background with $2c 2j 2 \gamma$ could have a similar cross section (5.8 pb) as that of $ p p \to 2b 2j 2\gamma$, but after the first two cuts and the mass window cut, the contribution of this type of background is only 8 fb or so, which is $4\%$ of that of $2b 2j 2 \gamma$ due to the fact that the mistagging rate is assumed to be $0.1$, in contrast to the tagging efficiency of the $b$ jet which is assumed to be $0.6$. Therefore, due to these two cuts, we simply omit the background events from the processes $ p p \to 4 j 2 \gamma$ and $ p p \to 2c 2j 2 \gamma$ in the following analysis.

All the preselection cuts are summarized in Table \ref{tab:table0}. After these cuts, the numbers of events are listed in Table \ref{tab:table1}.  The results given in Table \ref{tab:table1} explicitly demonstrate that the background events are so huge that the observation of triple-Higgs production is very challenging if no more analysis is conducted.

\begin{center}
\begin{table}
  \begin{center}
  \begin{tabular}{|c|c|}
  \hline
Preselection cuts & Description  \\ \hline
1 &  Number of tagged $b$-jets $n_b \geq 2$ and $P_t(j)>30$ GeV with $ 4 \leq n_j \leq 5$ \\ 
  \hline 
2 & Number of photons $n_\gamma=2$ with $P_t(\gamma)>30$ GeV  \\
  \hline
3 &   Number of leptons $n_l=0$ \\
  \hline
4 &   Missing energy cut MET$<50$ GeV    \\
  \hline
  \end{tabular}
  \end{center}
  \caption{\label{tab:table0}The preselection cuts in our analysis.}
\end{table}
\end{center}
\begin{center}
\begin{table}
  \begin{center}
  \begin{tabular}{|c|c|c|c|c|}
  \hline
                     &  $\sigma\times BR$ (fb)          &  K factors                          &  Events after preselection cuts  \\
  \hline
  Signal             &  $9.5\times 10^{-3}$             &  $2.0$  &  $50$      \\
  \hline
  $b\bar{b}jj\gamma\gamma$  &  $1.9 \times 10^{2} $                  &  $1.0$                              &  $2.3\times 10^{5}$    \\
  \hline
  $H(\gamma\gamma)t\bar{t}$  &  $77$                  &  $1.2$          &  $2.2\times 10^{4}$    \\
  \hline
  $S/B$                           & \multicolumn{3}{c|} {$1.9\times 10^{-4}$}                                          \\
  \hline
  $S/\sqrt{S+B}$                  & \multicolumn{3}{c|} {$9.8\times 10^{-2}$}                                          \\
  \hline
  \end{tabular}
  \end{center}
  \caption{\label{tab:table1}The total cross section and the number of events after preselection. Here, the total integrated luminosity is $30$ ab$^{-1}$. To appreciate the efficiency of each cut, the values of $S/B$ and $S/\sqrt{S+B}$ are provided. For the signal and $H(\gamma\gamma)t\bar{t}$ background, we adopt $K$ factors of $2.0$~\cite{Papaefstathiou:2015paa} and $1.2$~\cite{Dawson:2003zu}, respectively. The $K$ factor for the $b\bar b jj \gamma\gamma$ background is not shown in the literature. We take a representative value of 1.0. Discussions on its estimated value and its impacts on our results are presented in the Sec. \uppercase\expandafter{\romannumeral6}.}
\end{table}
\end{center}

\begin{figure*}[!htbp]
  \centering
\subfigure[$\chi^2_H$ in the Higgs reconstruction \label{fig3}]
 {\includegraphics[width=0.45\textwidth]{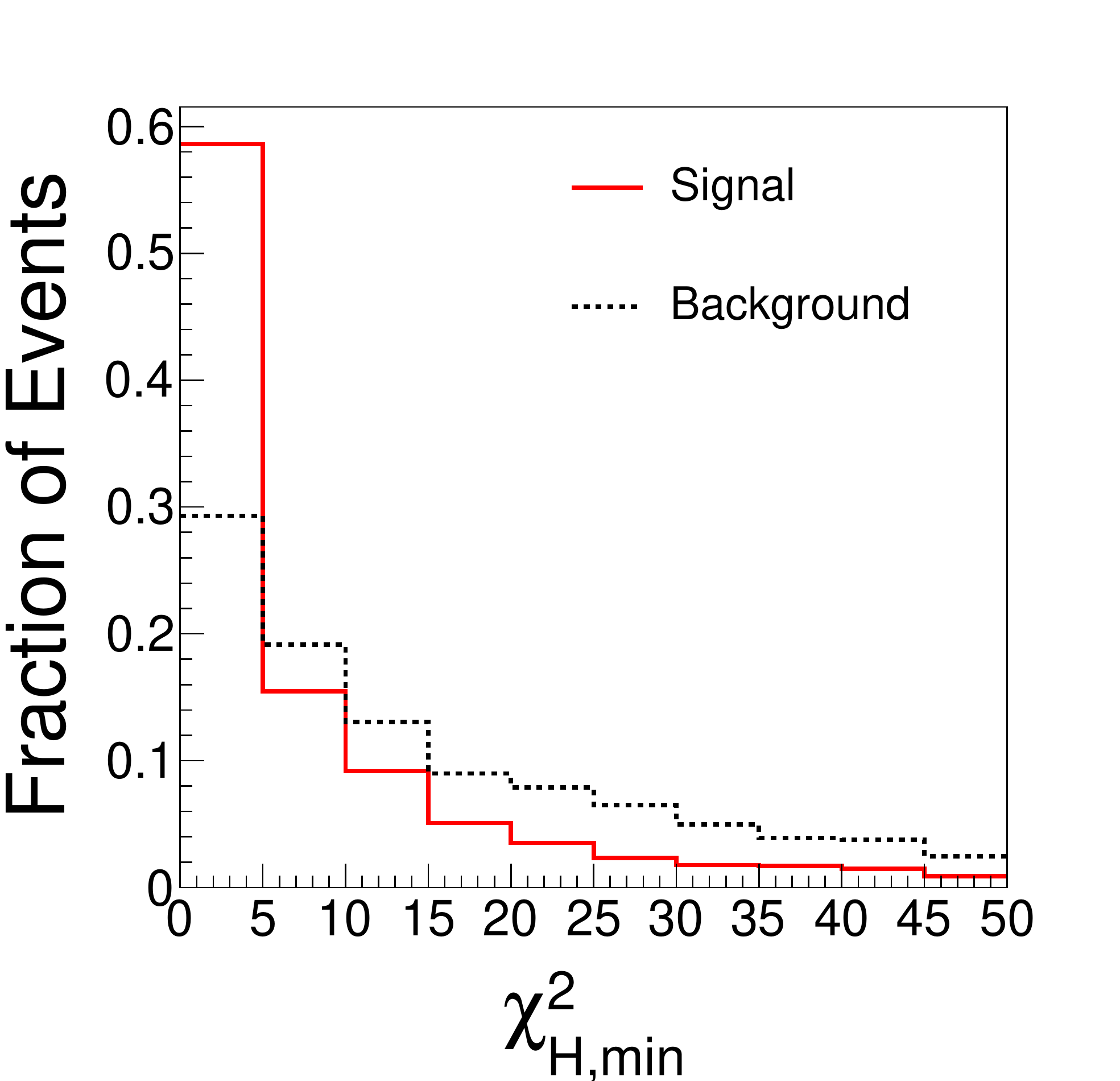}}
\subfigure[$\chi^2_t$ in the Higgs reconstruction \label{fig5}]
  {\includegraphics[width=0.45\textwidth]{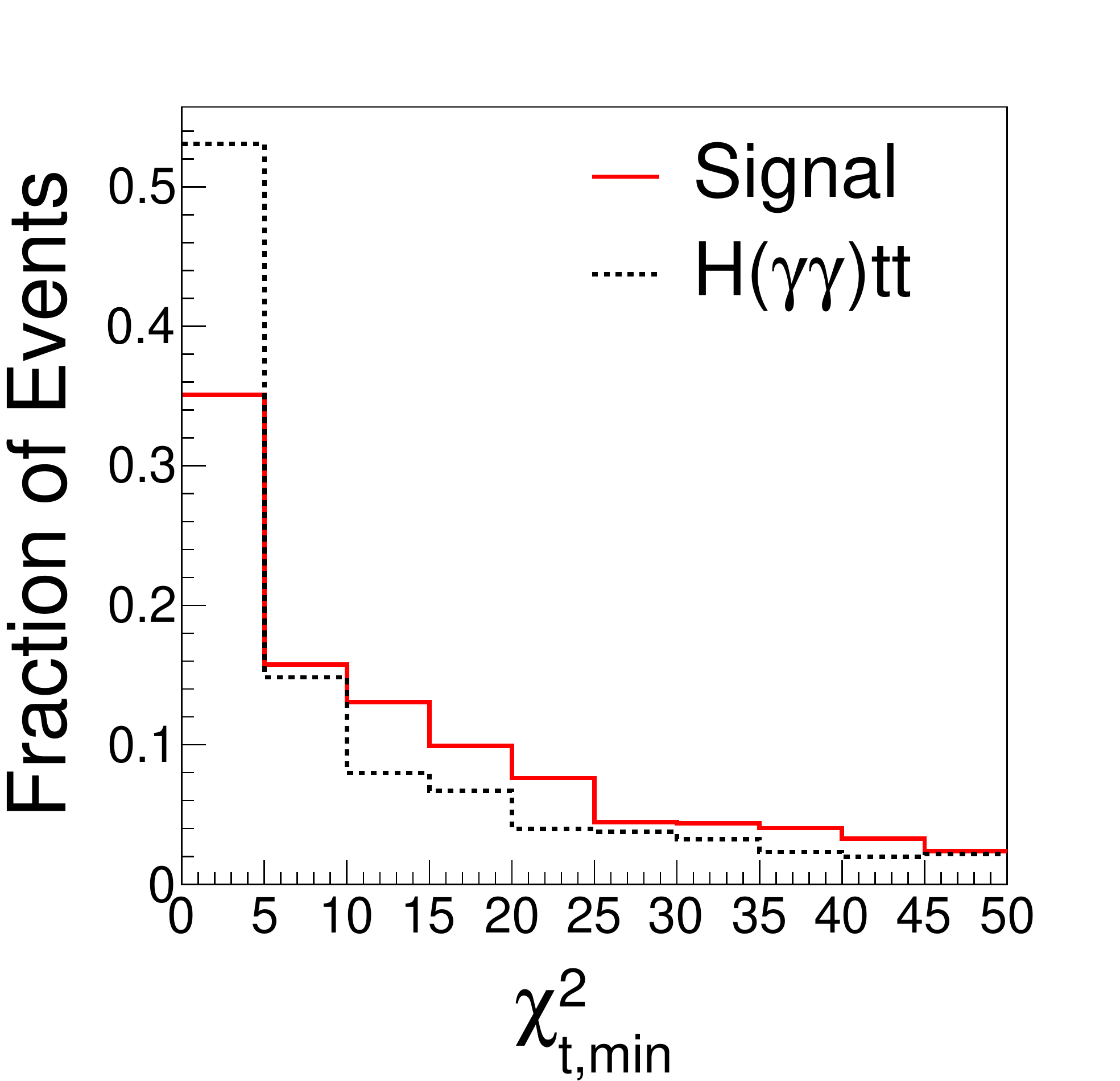}}\\
  \label{fig:chi2}
  \caption{The distributions of the minima of $\chi^2$ are shown.}
\end{figure*}

To further suppress the background by using the kinematics of the signal, we reconstruct the Higgs mass by introducing a $\chi^2$ method, where $\chi^2$ is defined as
\begin{eqnarray}
  \chi_H^2(m)=\frac{\left|M(j_1,j_2)-m\right|^2}{\sigma_j^2}+\frac{\left|M(j_3,j_4)-m\right|^2}{\sigma_j^2}+\frac{\left|M(\gamma,\gamma)-m\right|^2}{\sigma_\gamma^2}.
\end{eqnarray}
Here, $M(j_1,j_2)$ and $M(j_3,j_4)$ are the invariant masses of two pairs of hard jets of each event, and $\sigma_j=10$ GeV is the uncertainty of resolving two jets. $M(\gamma,\gamma)$ is the invariant mass of photons, and $\sigma_\gamma=\sqrt{2}$ GeV is the uncertainty of resolving a pair of photons. All combinations of pairing jets are considered, and the reconstruction mass $m^{rec}_H$ is chosen as the $m$ which minimizes $\chi_H^2$. The distribution of the minimum of $\chi_H^2$ is shown in Fig. \ref{fig3}. Here, we have combined $b\bar{b}jj\gamma\gamma$ events and $Ht\bar{t}$ events based on their weights in the total background. It can be seen that the background tends to have a large $\chi_{H,min}^2$, so we can introduce a cut $\chi_{H,min}^2<6.1$ to suppress the background.

Because the Higgs boson in a $Ht\bar{t}$ event decays to two photons, we noticed that the cut on $m_{\gamma\gamma}$ or $m^{rec}_H$ cannot suppress this type of background effectively. To veto such a type of background, we reconstruct the top by three jets. We use the reconstruction method described in Ref. \cite{Yang:2011jk}, where a $\chi^2$ for top reconstruction is
\begin{eqnarray}
  \chi_t^2=\frac{\left|M(j_1,j_2,j_3)-m_t\right|^2}{\sigma_t^2}+\frac{\left|M(j_1,j_2)-m_W\right|^2}{\sigma_W^2}.
\end{eqnarray}
Here $m_t=173$ GeV is the top mass, $m_W=80.4$ GeV is the $W$ mass, $\sigma_t=15$ GeV, and $\sigma_W=10$ GeV. The reconstructed top mass and $W$ mass are defined as $M^t_{rec}=M(j_1,j_2,j_3)$ and $M^W_{rec}=M(j_1,j_2)$ when $\chi_t^2$ is minimum. In the top reconstruction, all combinations of pairing jets are considered, and we require that  $M(j_1,j_2)$ does not include $b$ jets if only two jets are tagged. The distribution of the minimum of $\chi_t^2$ is shown in Fig. \ref{fig5}. 

\begin{figure}[htbp]
  \centering
  \subfigure{
  \label{Fig6.sub.1}\thesubfigure
  \includegraphics[width=0.4\textwidth]{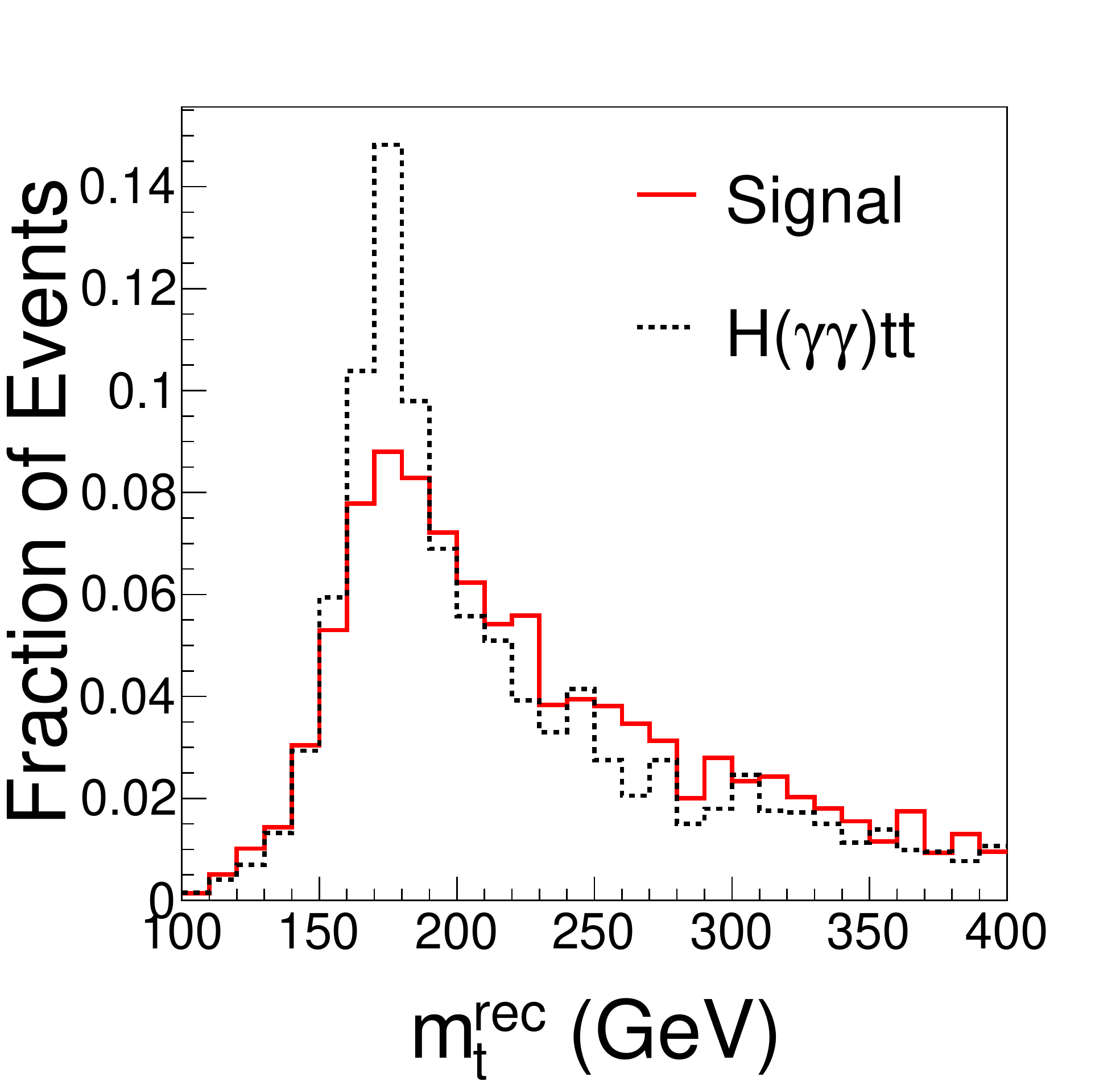}}
  \subfigure{
  \label{Fig6.sub.2}\thesubfigure
  \includegraphics[width=0.4\textwidth]{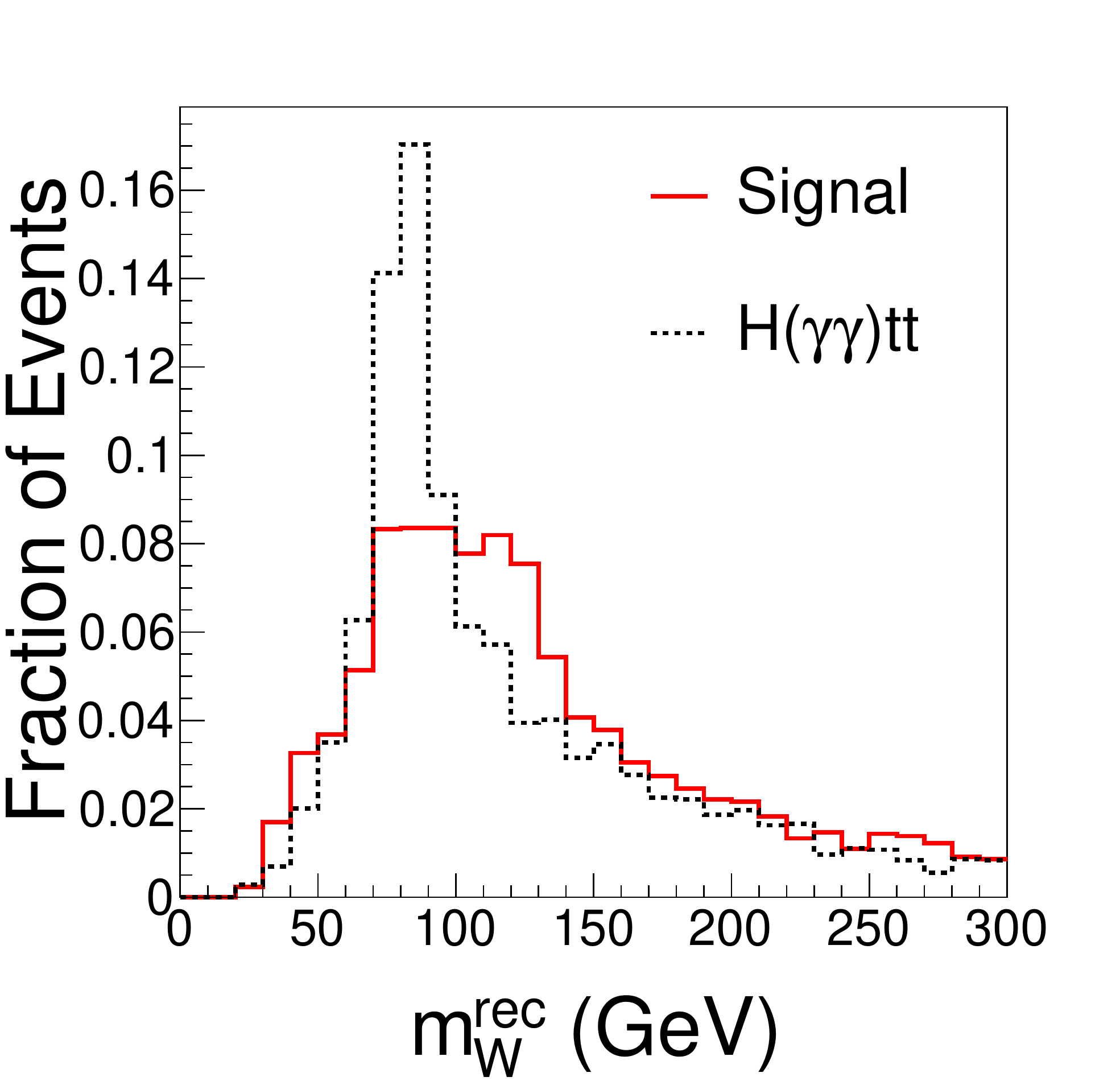}}
  \caption{The distributions of the (a) reconstructed top mass and (b) reconstructed $W$ mass.}\label{fig6}
\end{figure}

The reconstructed top and $W$ masses are shown in Fig. \ref{fig6}. There are peaks around $m_t^{rec}=173$ GeV and $m_W^{rec}=80$ GeV, both in the signal and backgrounds due to the constraint in the definition of $\chi^2_t$. However, there is another peak around $m_W^{rec}=126$ GeV in Fig. \ref{Fig6.sub.2}, which indicates that these jets have decayed from the Higgs boson. 

\begin{figure}[htbp]
  \centering
  \subfigure{
  \label{Fig4.sub.1}\thesubfigure
  \includegraphics[width=0.4\textwidth]{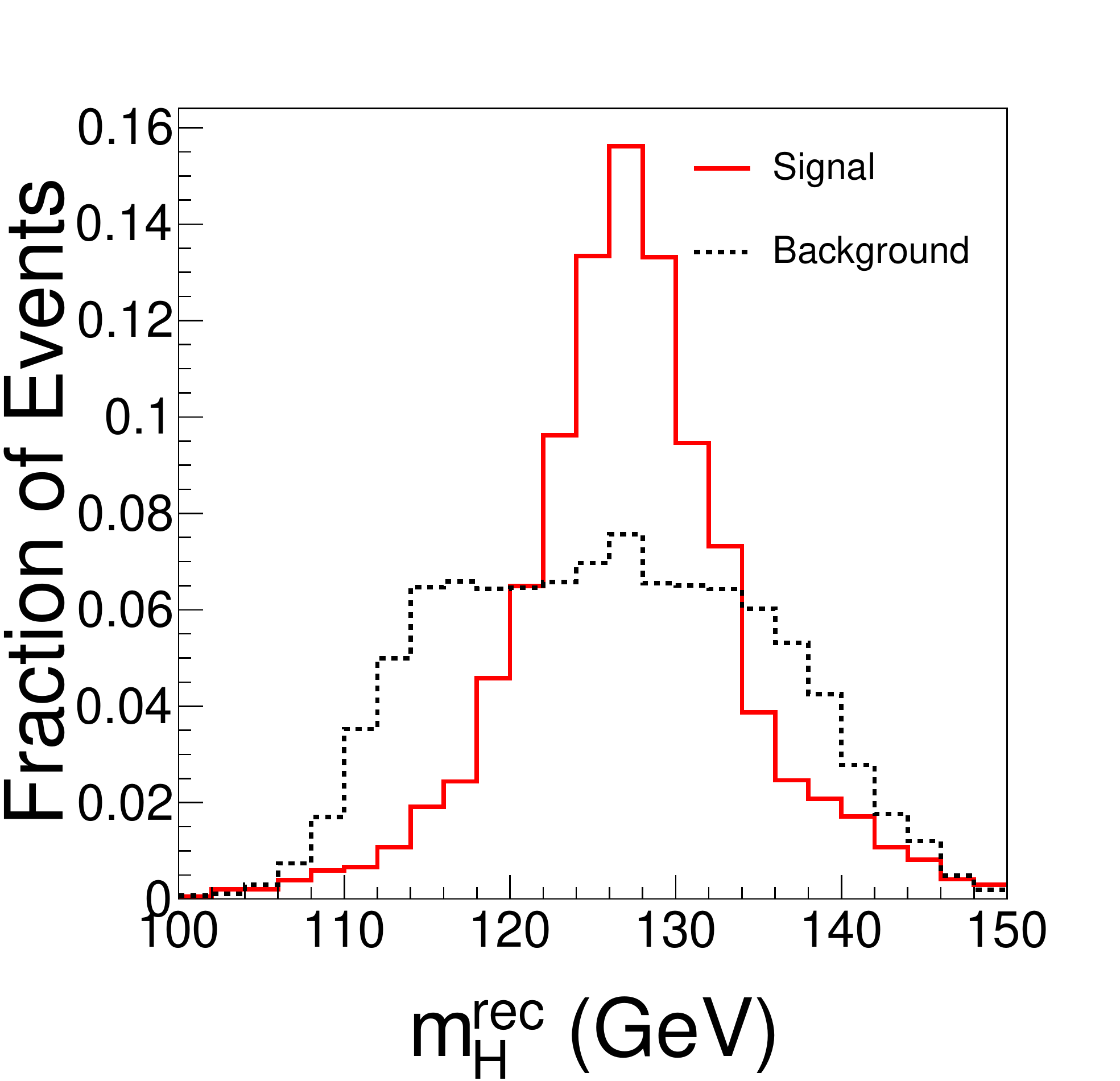}}
  \subfigure{
  \label{Fig4.sub.2}\thesubfigure
  \includegraphics[width=0.4\textwidth]{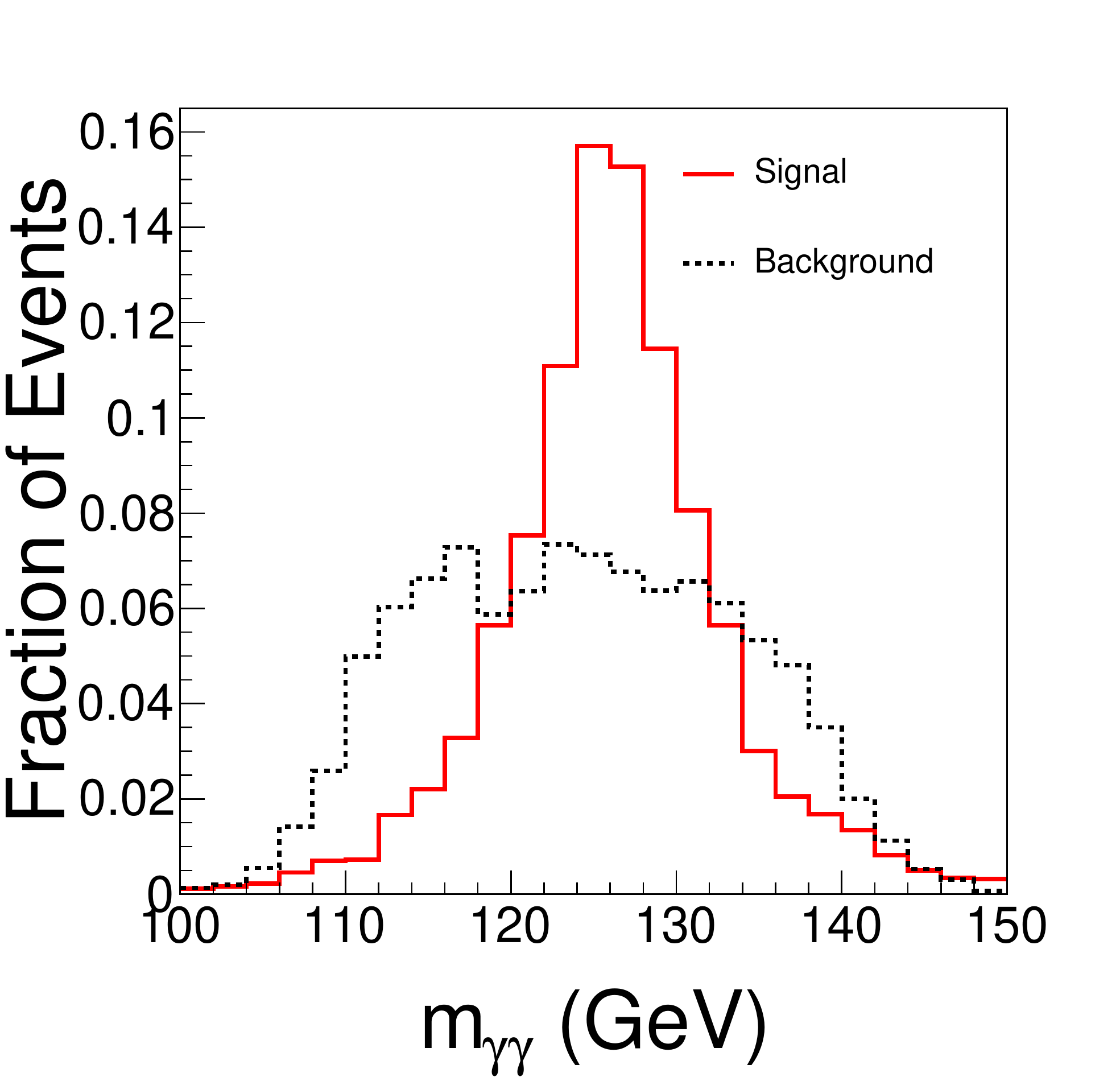}}
  \subfigure{
  \label{Fig4.sub.3}\thesubfigure
  \includegraphics[width=0.4\textwidth]{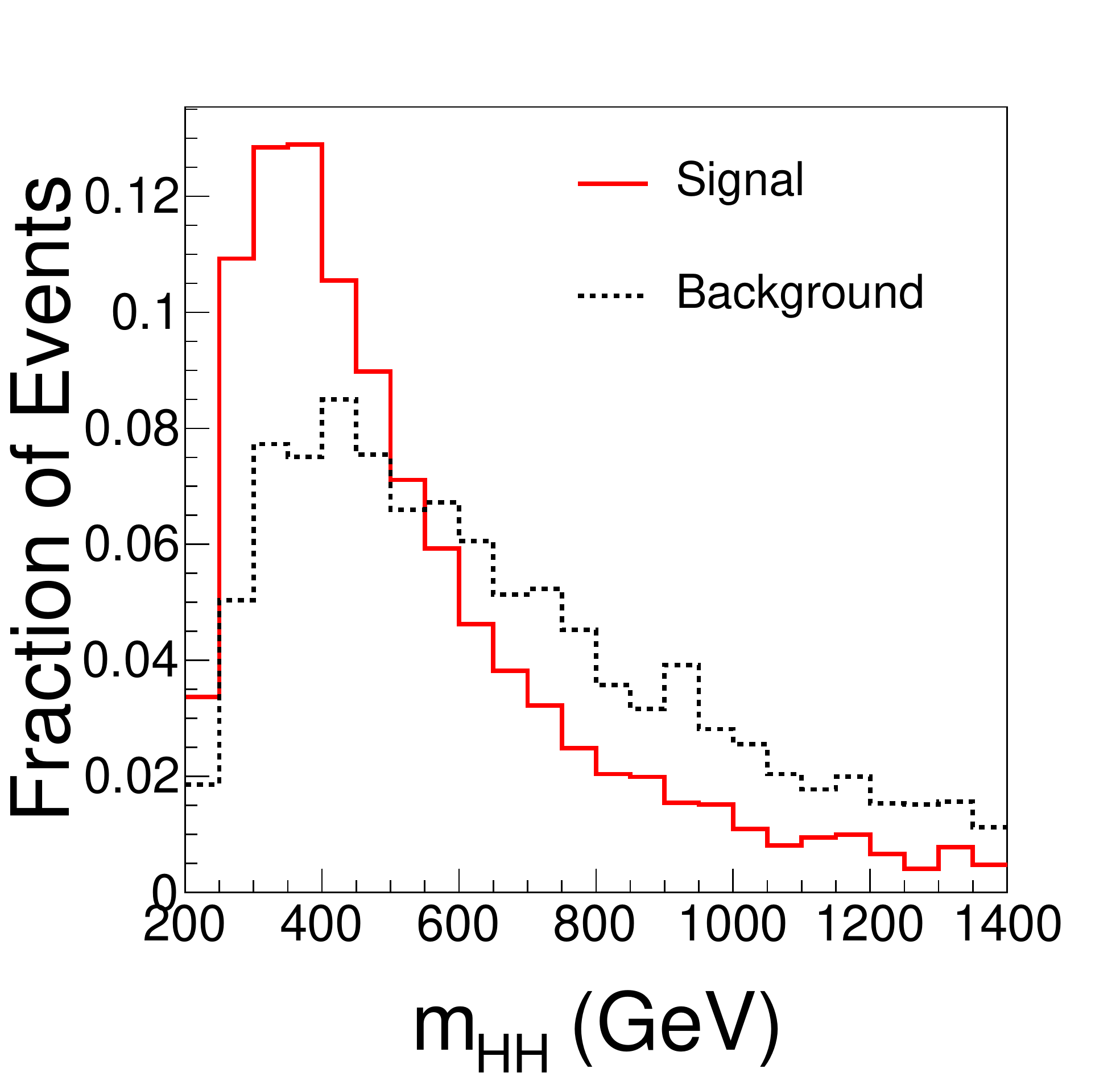}}
  \subfigure{
  \label{Fig4.sub.4}\thesubfigure
  \includegraphics[width=0.4\textwidth]{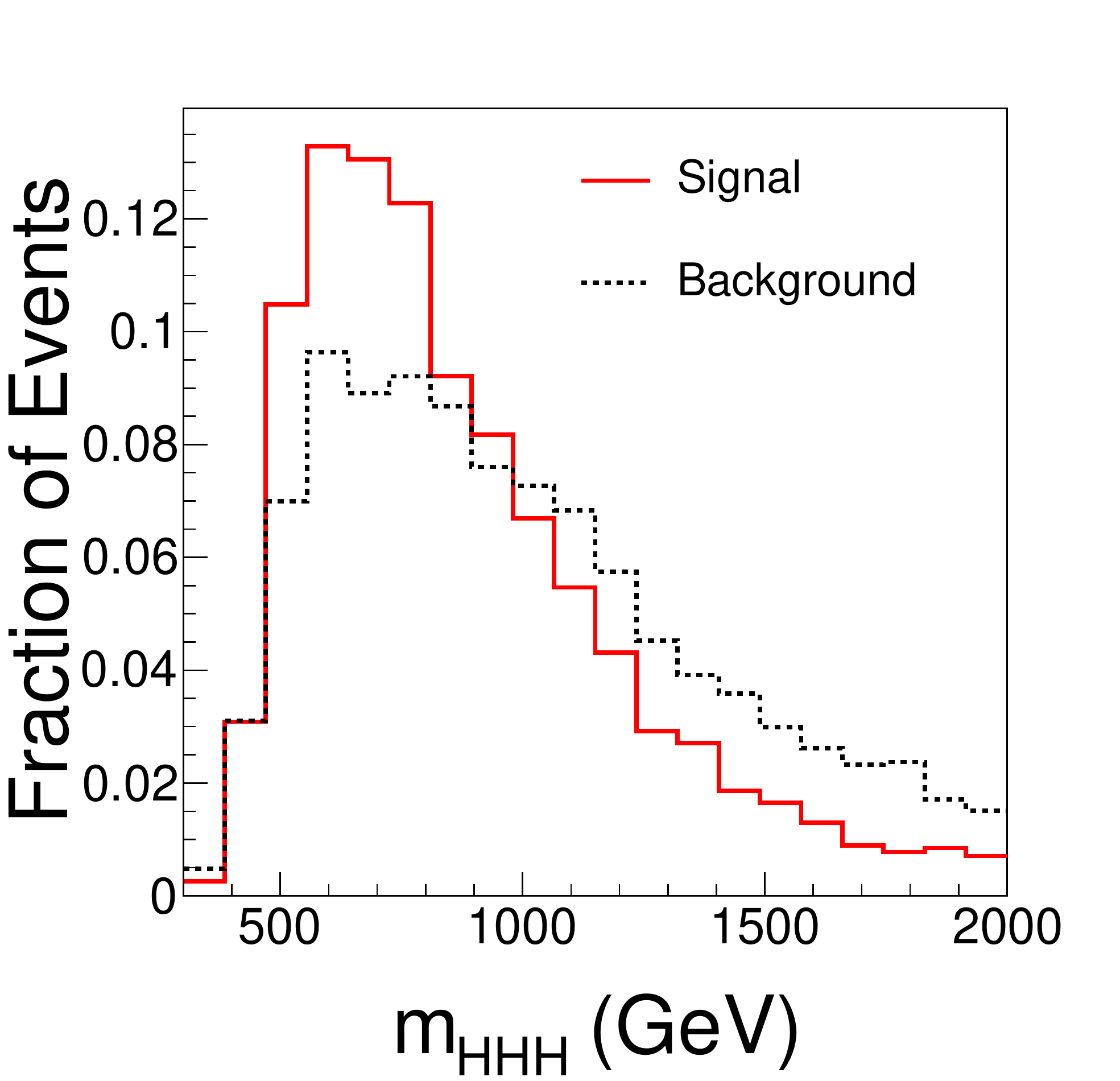}}
  \caption{The distributions of the (a) recontructed Higgs mass, (b) invariant mass of two photons, (c) invariant mass of the hardronic Higgs, and (d) total invariant mass of three Higgses.}\label{fig4}
\end{figure}

We are interested in three invariant-mass variables: the reconstructed Higgs mass ($m^{rec}_H$), the invariant mass of the hadronic Higgs bosons ($m_{HH}$), and the total invariant mass of Higgs bosons ($m_{HHH}$). They can be extracted after the reconstruction of  Higgs bosons. The distributions of these observables are shown in Fig. \ref{fig4}. In Fig. \ref{Fig4.sub.1}, there is a peak around $m^{rec}_H=126$ GeV of signal, but the distribution of the background is flat at the region $100$ GeV$<M_H<150$ GeV, which is consistent with the cuts we imposed at the generator level. After taking the resolution power of photons into consideration, we introduce a reconstructed mass cut $|m^{rec}_H-126$ GeV$|<5$ GeV. Fig. \ref{Fig4.sub.2} shows the distribution of the invariant mass of photons. The decay width effect of Higgs boson is not considered in our analysis, so the broadening of the peak in the invariant mass $m_{\gamma\gamma}$ is attributed to the detector effects. The invariant mass of photons gives a strong constraint on $m^{rec}_H$, so a peak can be observed in Fig. \ref{Fig4.sub.1}. The peak of Higgs boson mass is reconstructed from a diphoton rather than photons from QCD, as shown in Fig. \ref{Fig4.sub.1} and Fig. \ref{Fig4.sub.2}. The invariant mass of two Higgs bosons which decay to $b\bar{b}b\bar{b}$, and total invariant mass of triple-Higgs, respectively, are shown in Fig. \ref{Fig4.sub.3} and Fig. \ref{Fig4.sub.4}. Because of the detector effects, the distributions of these observables are broadened when compared with those at parton-level ones given in Figs. \ref{Fig1.sub.1} and Fig. \ref{Fig1.sub.2}. 

All cuts we introduced are concluded in Table \ref{tab:table2}. This result shows that the cuts we have introduced can enhance $S/B$ by almost $1$ order of magnitude but cannot improve  $S/\sqrt{S+B}$ too much. The smallness of the signal cross section and the detector effects prevent effective background suppression.
\begin{center}
\begin{table}
  \begin{center}
  \begin{tabular}{|c|c|c|c|}
  \hline
                        &  Signal          &  $b\bar{b}jj\gamma\gamma$   &  $Ht\bar{t}$       \\ 
  \hline
  Preselection          &  $50$                    &  $2.3\times 10^{5}$               &  $2.2\times 10^{4}$ \\ 
  \hline
  $\chi_{H,min}^2<6.1$  &  $26$                    &  $4.6\times 10^{4}$               &  $9.9\times 10^{3}$ \\ 
  \hline
  $|m^{rec}_H-126$ GeV$|<5.1$ GeV  &  $20$         &  $1.7\times 10^{4}$               &  $7.0\times 10^{3}$ \\ 
  \hline
  $S/B$                           & \multicolumn{3}{c|} {$8.3\times 10^{-4}$}   \\
  \hline
  $S/\sqrt{S+B}$                  & \multicolumn{3}{c|} {$0.13$}            \\
  \hline
  \end{tabular}
  \end{center}
  \caption{\label{tab:table2}The efficiency of the cuts are demonstrated. Here, the total integrated luminosity is $30$ ab$^{-1}$. To appreciate the efficiency of each cut, the values of $S/B$ and $S/\sqrt{S+B}$ are provided.}
\end{table}
\end{center}

\subsection{Multivariate analysis}
We apply two multivariate analysis approaches, 1) the boost decision tree (BDT) and 2) multilayer perceptron (MLP) neural network, to utilize the correlation of observables in the signal to further suppress backgrounds. In this case, we only consider the events with four jets exactly and do not introduce any cuts on MET. The observables $P_t(j_i)$, $P_t(\gamma_i)$, $\eta(j_i)$, and $\eta(\gamma_i)$ are considered, where $i=1,2,3,4$ for jets and $i=1,2$ for photons. In addition, the observables we discussed above (MET, $\chi_{H,min}$, $\chi^2_{t,min}$, $m^{rec}_H$, $m_{\gamma\gamma}$, $m_{HH}$, $m_{HHH}$, $m_t^{rec}$, and $m_W^{rec}$) are also used. 

\begin{figure}[htbp]
  \centering
  \subfigure{
  \label{Fig7.sub.1}\thesubfigure
  \includegraphics[width=0.4\textwidth]{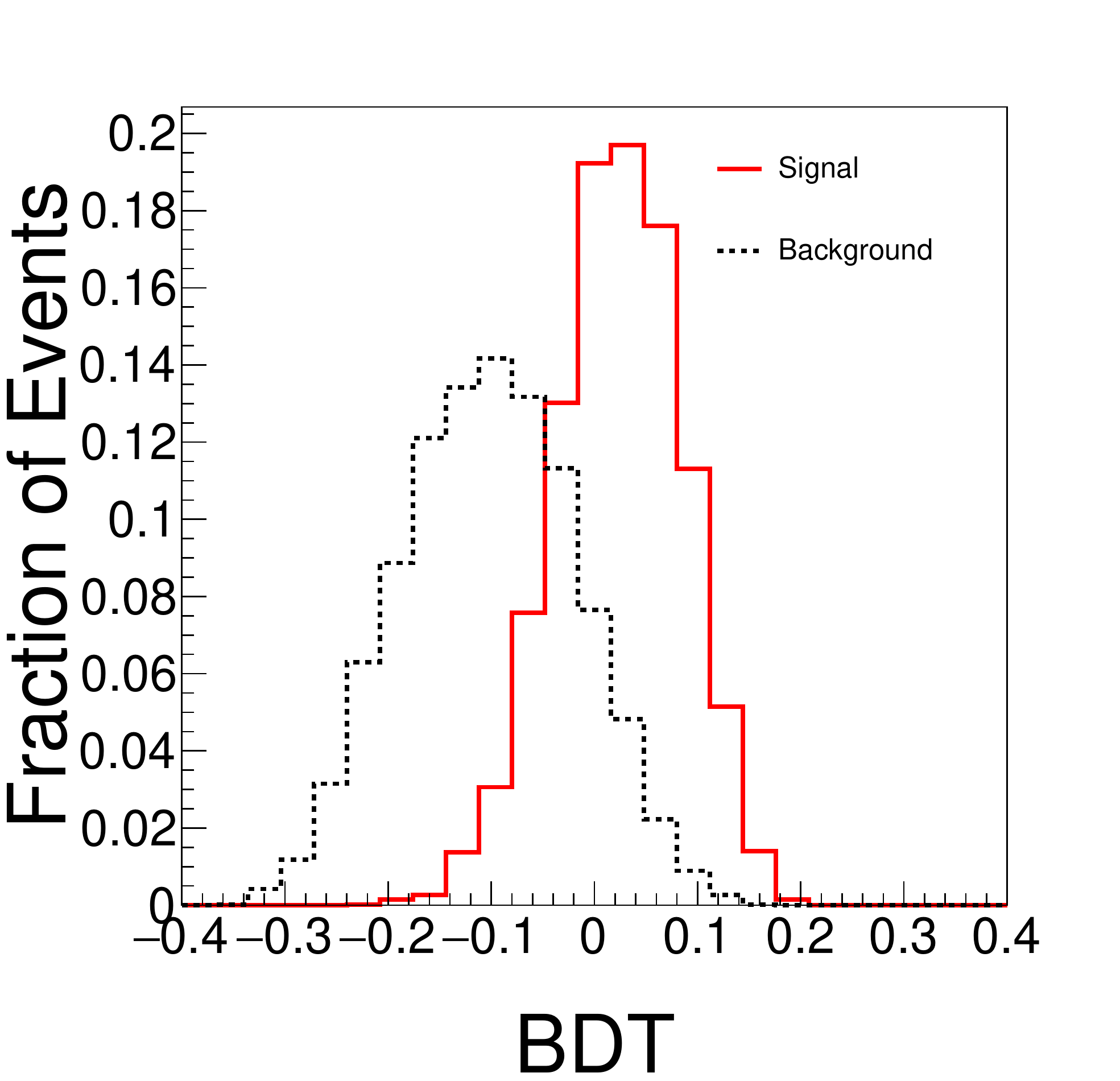}}
  \subfigure{
  \label{Fig7.sub.2}\thesubfigure
  \includegraphics[width=0.4\textwidth]{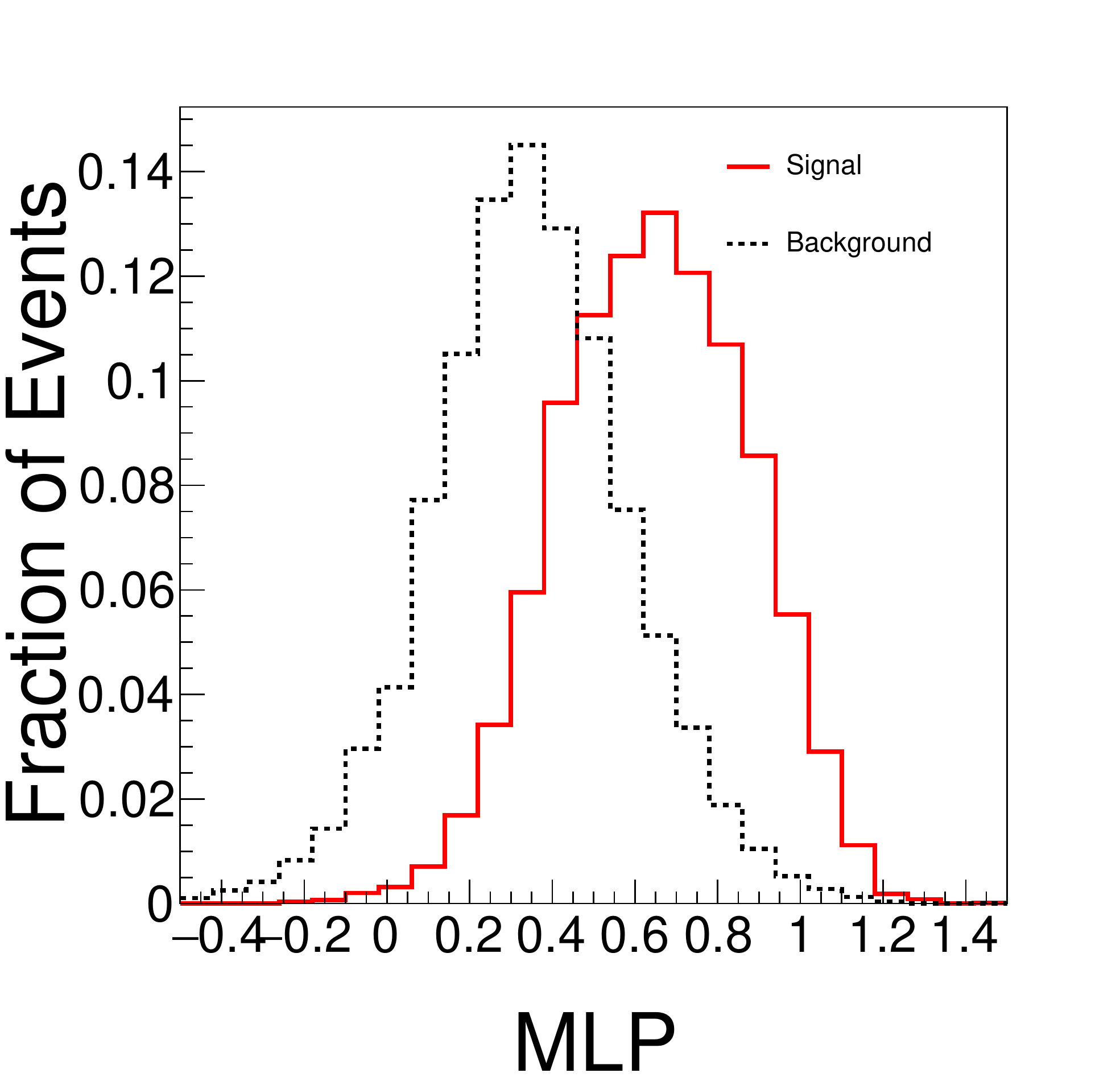}}
  \caption{The response of the discriminants to the signal and background in two multivariate analyses, (a) the BDT method and (b) the MLP neural network method.}\label{fig7}
\end{figure}

The results are presented in Fig. \ref{fig7}, and the efficiencies are summarized  in Table \ref{tab:table3}. The BDT method can increase the value $S/\sqrt{S+B}$ to $0.20$, which can be much better than that of the simple cut method. But it is still far from the discovery of the triple-Higgs signal. 

To observe the triple-Higgs signal of the SM at the 5$\sigma$ level, a much larger integrated luminosity is necessary.
Table \ref{tab:table4} shows the values of $S/\sqrt{S+B}$ at different integrated luminosity. There, we scale up the integrated luminosity for both the signal and background. From the table, we see that the integrated luminosity should be around $1.8\times 10^4$ ab$^{-1}$ if we want to discover the triple-Higgs production via the $b\bar{b}b\bar{b}\gamma\gamma$ mode at a 100 TeV machine. If we want to extract the information of $\lambda_4$, we need an even larger luminosity, as we can see from Eq. (\ref{xsl4}), where the coefficient $B^\prime$ of $\lambda_4$ is only one-eighth of $C^\prime$. This is indeed challenging when considering the realistic integrated luminosity for the future collider projects, as addressed in Ref. \cite{Hinchliffe:2015qma}.

\begin{table}
  \begin{center}
  \begin{tabular}{|c|c|c|c|}
  \hline
              &  Cuts-based method     &  BDT$>0.02$                   &  MLP$>0.51$  \\
  \hline
  Signal      &  $20$                 &  $34$                        &  $49$      \\
  \hline
  Background  &  $2.4\times{10^4}$     &  $2.8\times{10^4}$             &  $9.9\times{10^4}$    \\
  \hline
  $S/B$              &  $8.3\times{10^{-4}}$  &  $1.2\times{10^{-3}}$          &  $5.0\times{10^{-4}}$    \\
  \hline
  $S/\sqrt{S+B}$       &  $0.13$                &  $0.20$                       &  $0.16$    \\
  \hline
  \end{tabular}
  \end{center}
  \caption{\label{tab:table3}The number of events and the significances of the BDT and MLP neural network method are demonstrated. Here, the total integrated luminosity is $30$ ab$^{-1}$.}
\end{table}

\begin{table}
  \begin{center}
  \begin{tabular}{|c|c|c|c|c|}
  \hline
  Integrated luminosity (ab$^{-1}$) &  $30$       &  $300$               &  $3000$  &  $1.83\times 10^4$ \\
  \hline
  $S/\sqrt{S+B}$                    &  $0.2$     &  $0.6$               &  $2.0$   &  $5.0$ \\
  \hline
  \end{tabular}
  \end{center}
  \caption{\label{tab:table4}The values of $S/\sqrt{S+B}$ with BDT$>0.02$ at different assumed integrated luminosities are displayed.}
\end{table}

\section{The sensitivity to quartic coupling}

\begin{figure}[htbp]
  \centering
    \vspace{0.2cm}
  \subfigure{
  \label{Feynman.sub.1}
  \includegraphics[width=0.22\textwidth]{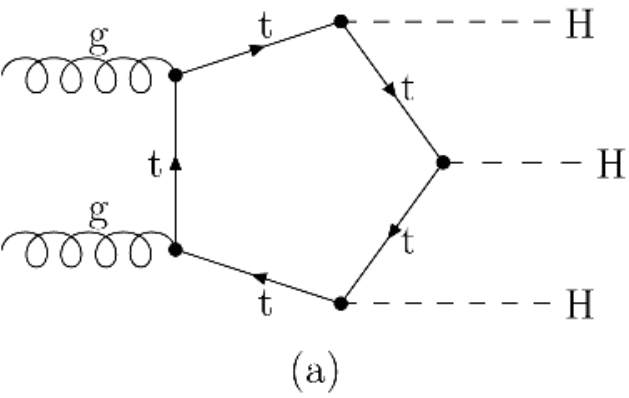}}
  \subfigure{
  \label{Feynman.sub.2}
  \includegraphics[width=0.22\textwidth]{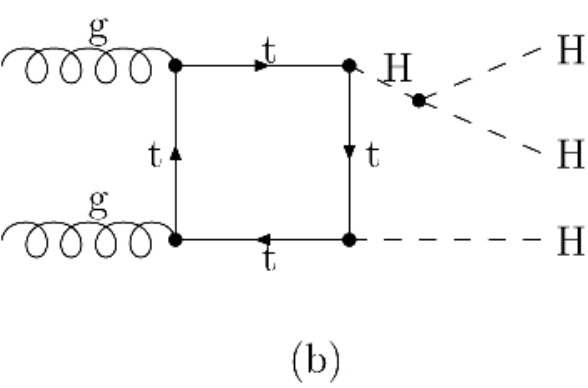}}
  \subfigure{
  \label{Feynman.sub.3}
  \includegraphics[width=0.22\textwidth]{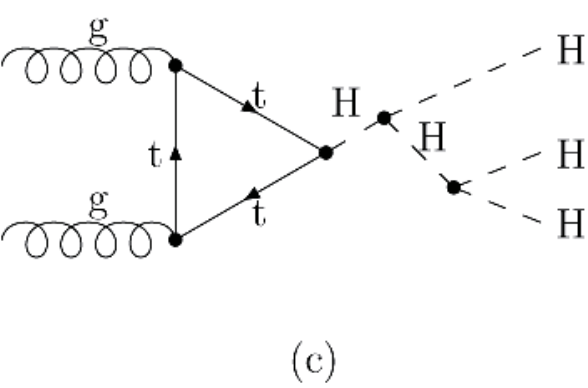}}
  \subfigure{
  \label{Feynman.sub.4}
  \includegraphics[width=0.22\textwidth]{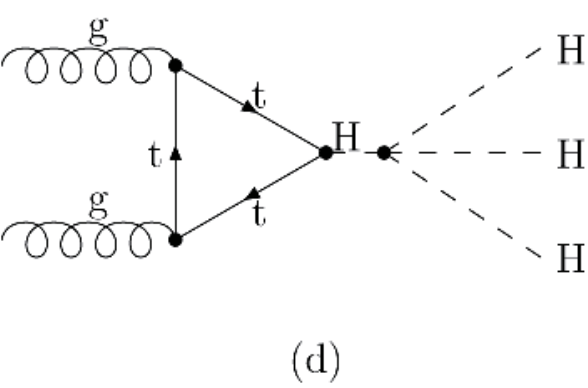}}
  \vspace{-0.1cm}
  \caption{The example Feynman diagrams of the process $gg\to HHH$ in the SM.}\label{feynman}
\end{figure}

It is well known that the process $gg\rightarrow HHH$ includes four kinds of Feynman diagrams, as shown in Fig. \ref{feynman}. They are as follows: three Higgs bosons are produced by a pentagon quark loop [Fig. \ref{Feynman.sub.1}], two Higgs bosons are produced by a box quark loop with a subsequent decay via trilinear coupling [Fig. \ref{Feynman.sub.2}], a Higgs boson is produced by a triangle quark loop and then decay to three Higgses through two trilinear vertices [Fig. \ref{Feynman.sub.3}], and the triangle quark loop produce a Higgs boson which decays to three Higgs bosons through quartic coupling [Fig. \ref{Feynman.sub.4}]. Only the last kind of diagram involves the quartic coupling. 

To explore the dependence of the cross section of the process $gg \to  HHH$ upon the parameters $\lambda_3$ and $\lambda_4$, we can use the Feynman diagrams as a guide and can parametrize the cross section in the form
\bea
\sigma(\lambda_3,\lambda_4) &= A \lambda_4^2 + \left( B \lambda_3^2 + C \lambda_3 + D \right) \lambda_4   \nonumber \\
&+E \lambda_3^4 +  F \lambda_3^3 + G  \lambda_3^2 + H  \lambda_3 + I\,, \label{xsl34}
\eea
where the coefficients $A$--$I$ can be determined by choosing a certain number of cross section values which we are able to determined by a set of input pairs of $(\lambda_3, \lambda_4)$. It should be pointed out that in this formula we have not included the NLO corrections. We have chosen $21$ cross section values in total by using our codes and determined the fitted coefficients $A$--$I$, which are tabulated below.
\begin{table}[th]
\begin{center}%
\begin{tabular}{|c|c|c|c|c|c|c|c|c|}
\hline
$ A$ & $B $& $C$ & $D$& $E$ & $F$ & $G$ & $H$ & $I$ \\ \hline
$5.28 \times 10^{-2}$  & $0.14$ & $-0.76 $& $0.15 $ &$2.28 \times 10^{-2}$&$-5.36 \times 10^{-2}$& $3.11 $&  $-14.57$ & $15.36$ \\ \hline
\end{tabular}
\end{center}
\caption{The fitting coefficients of Eq.~(\ref{xsl34}). \label{table1No}}%
\end{table}

\begin{figure*}[!htbp]
\subfigure[ ]
{\includegraphics[width=.45 \textwidth]{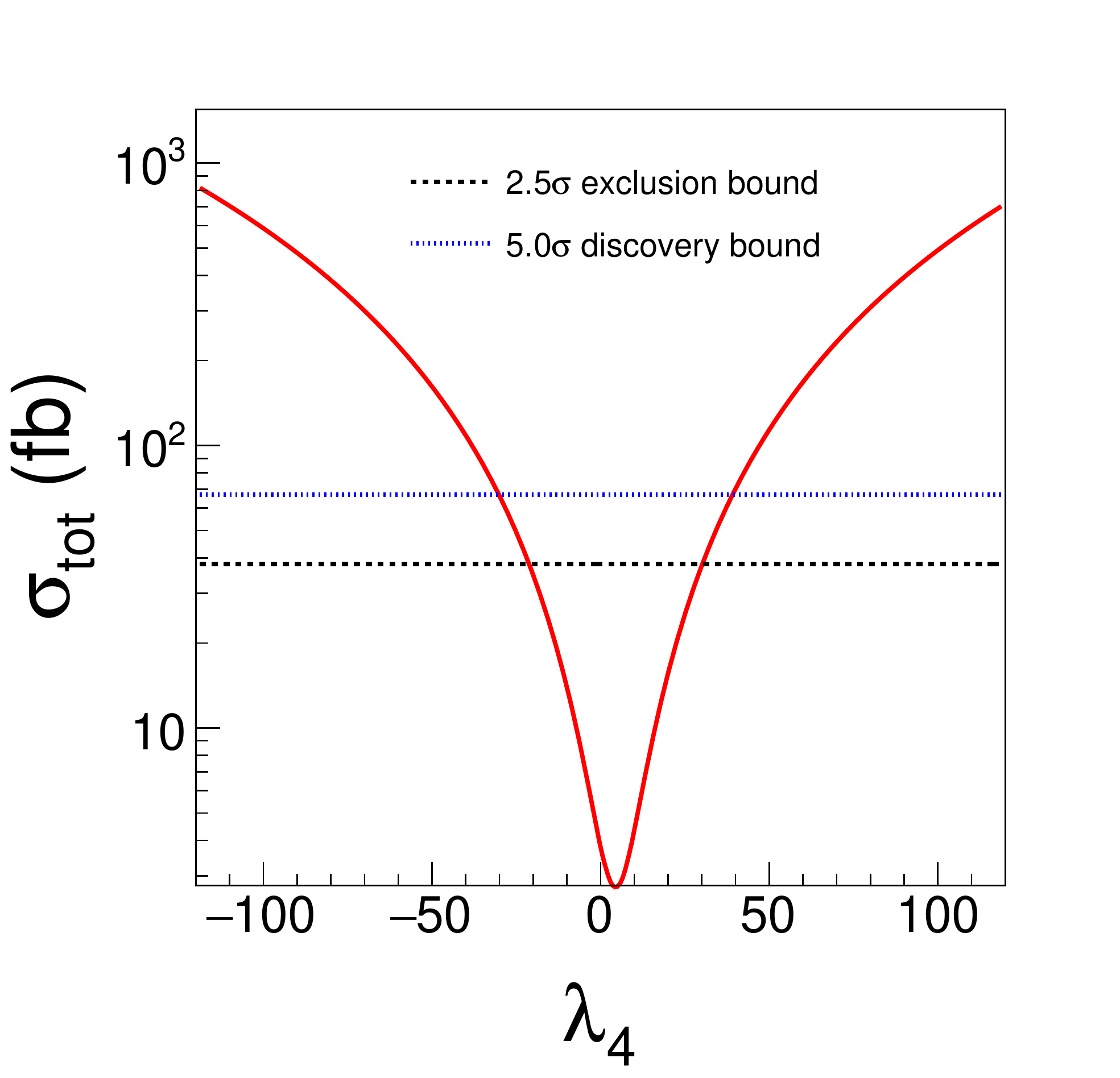} \label{l4xs}}
\subfigure[ ]
{\includegraphics[width=.45 \textwidth]{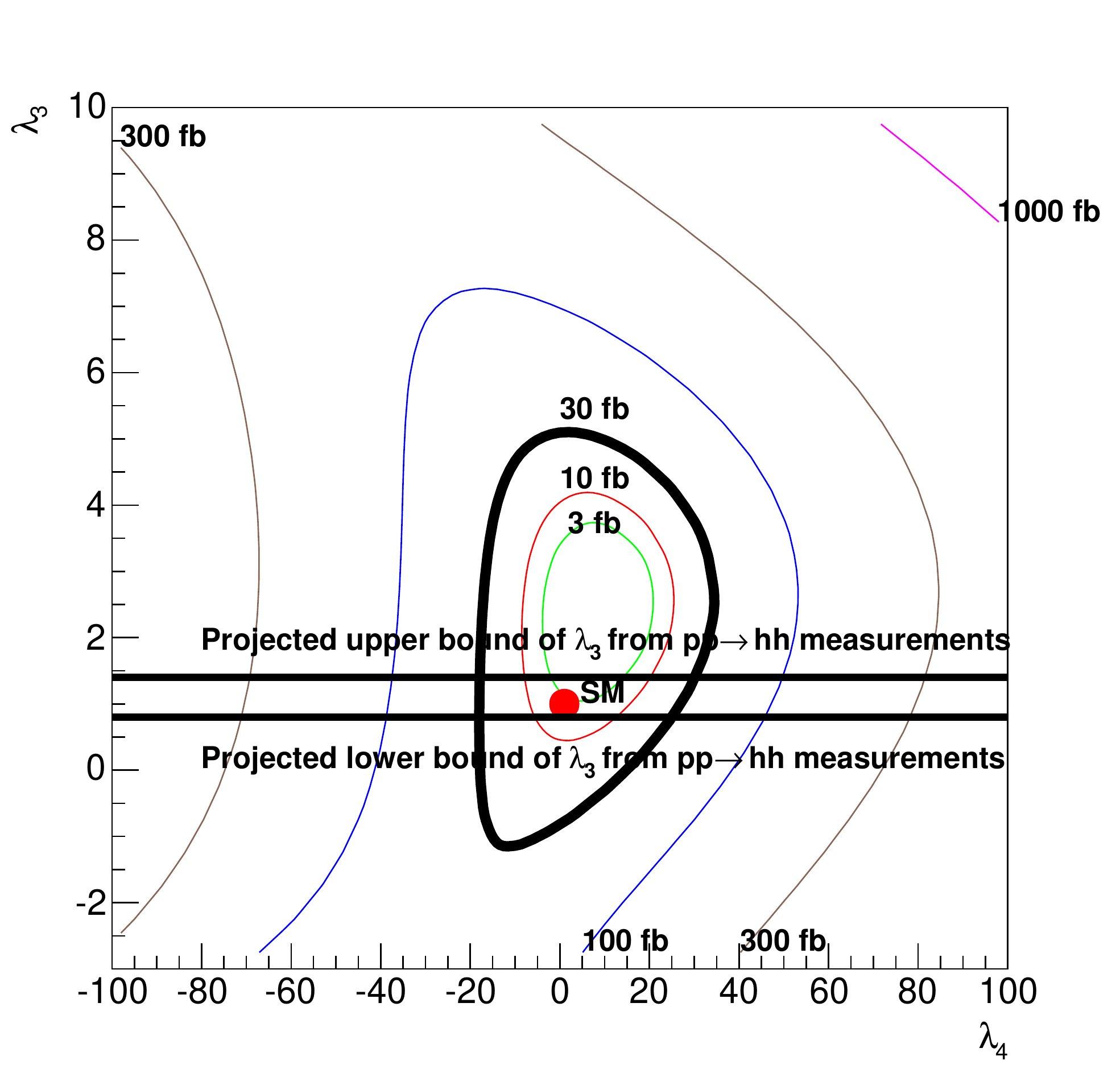} \label{l3l4}} \\
\caption{(a) The fitted cross section when $\lambda_3=1$; (b) the feasibility contours of $\sigma(pp\to hhh)$ in the $\lambda_4-\lambda_3$ plane.}
\end{figure*}

From the fitted coefficients given in Table \ref{table1No}, a few comments are in order:
\begin{enumerate}
\item The largest three are $G$, $H$, and $I$. $I$ is the contribution of the pentagon diagram. The term proportional to $G$ is the contribution of box diagrams. And the term proportional to $H$ corresponds to the interference between the pentagon diagram and box diagrams.

\item The sign of $H$ is opposite those of $G$ and $I$. Consequently, the total cross section could be sensitive to the sign of $\lambda_3$; when $\lambda_3$ is positive, it corresponds to a destructive interference, and when $\lambda_3$ is negative, it corresponds to a constructive interference. It is the former case for the SM.

\item The coefficients $A$, $E$, and $F$, are of order ($10^{-2}$) and are proportional to $\lambda_4^2$, $\lambda_3^4$, and $\lambda_3^3$, respectively. These three terms can only be large when $\lambda_4$ and $\lambda_3$ are significant.

\item The interference between the triangle and pentagon/box/triangle diagrams are proportional to $B$, $C$, and $D$. 
It is of the order $\mathcal O(10^{-1})$. It should be noticed that the sign of $C$ is different from those of $B$ and $D$, which indicates that a destructive interference occurs in 
the SM.

\item When $\lambda_3$ is fixed to the SM value ,
i.e., $\lambda_3=1$, the cross section can be simply parametrized as
\bea
\sigma(\lambda_4) &= A \lambda_4^2 + B^\prime \lambda_4 + C^\prime \,. \label{xsl4}
\eea
We find that $B^\prime=-0.47$ and $C^\prime=3.82$, which is consistent with the formula given in Eq. (\ref{xsl34}). The fitted cross section is shown in Fig. \ref{l4xs}. 
It shows good agreement in our numerical results. The minimal value of the cross section happens when $\lambda_4=4.46$ and the corresponding cross section is 2.77 fb.

\end{enumerate}

By using the fitted cross section given in Eq. (\ref{xsl34}) and combining it with our feasibility analysis given in the section above, we explored the projected sensitivity of a 100 TeV collider project to both $\lambda_3$ and $\lambda_4$ from the measurement of $pp \to hhh$ via the $4b$ and 2$\gamma$ final states. The result is demonstrated in Fig. \ref{l3l4}.

In Fig. \ref{l3l4}, we show six contours of the cross section which correspond to $1000$ fb (pink), $300$ fb (yellow), $100$ fb (blue), $30$ fb (black), $10$ fb (red), and $3$ fb (green), respectively. It should be noticed that the $K$ factors of the signal are not included in this plot. If they were included, the results could be better.

Among them, we estimate that the contour with $30$ fb is the minimal required cross section for the discovery, which is depicted by a dark line; the contour with $3$ fb is  depicted by a green line, which is close to the cross section of the SM. In the plot, the big red spot denotes the value of the SM. It is worth mentioning that to reach 30 fb the value of $\lambda_4$ is so large that the perturbativity and the perturbative unitarity are violated.

From the contour with 30 fb, we can read that to discover $gg\to HHH$ the parameter $\lambda_3$ should be confined to the range $[-1,5]$ and $\lambda_4$ should be confined to the range $[-20,30]$.

For the purpose of comparison, we also depict the projected upper and lower bounds of $\lambda_3$ from the measurements of di-Higgs production from the final states $b {\bar b} \gamma \gamma$ \cite{Yao:2013ika} and $3 \ell 2j + \missE$ \cite{Li:2015yia}, which could narrow the value of $\lambda_3$ down to $1^{+0.4}_{-0.2}$ due to its larger production rate. 

\section{Triple-Higgs Production in the Higgs Singlet Model}
Although the Higgs boson has been discovered, the direct measurement of Higgs self-couplings is still 
under confirmation. Exploring the shape of the electroweak (EW) Higgs potential is extremely important and could serve as a window to new physics. Probing Higgs self-couplings can either confirm the SM or discover new physics, which is a no-lose theorem.  

In addition, the matter and antimatter asymmetry has been one of the most fundamental questions in particle physics. A very promising solution is baryogenesis, which requires three criteria to explain the generation of baryon asymmetry observed in the present Universe: 1) baryon number violation, 2) $C$ and $CP$ violations, and 3) departure from thermal equilibrium. 
In the SM, the $CP$-violation phase is not big enough. Furthermore, even if the $CP$-violation phase is sufficiently large,
for a Higgs with mass at 125--126 GeV, the first-order phase transition is not strong enough. 
This gives us a strong motivation to introduce new physics.

We have learned that the production rate of triple-Higgs events is small in the SM, but it can be enhanced dramatically in a new physics model. One simple extension is adding a real scalar singlet to the SM Higgs sector \cite{Pruna:2013bma,Barger:2007im,O'Connell:2006wi,Profumo:2014opa,Chen:2014ask}. 
Moreover, in this model,  
it is straightforward to produce a strong first-order phase transition \cite{Ahriche:2007jp,Profumo:2007wc}. In particular, we find that there exists a part of parameter space where the quartic couplings play important roles.  Although the main discovery channels are still through $H_2 \to WW, ZZ$, and $t{\bar t}$ (which can either be used to determine the value of the mixing angle or put a constraint on it), triple-Higgs production can provide another opportunity to directly observe a new heavy scalar if $BR(H_2 \to H H H)$ is sizeable and thus open up the possibility of a precision measurement of the quartic couplings. 
Therefore, we propose a new channel in which a heavy singlet scalar is produced at resonance and decays into three 126 GeV Higgs bosons. We point out that in this part of parameter space the resonant di-Higgs production is highly suppressed, and the resonant triple Higgs production becomes an important channel to look for the new heavy singlet scalar. 

In the singlet+SM model, the Higgs potential can be parametrized as \cite{Chen:2014ask}
\begin{eqnarray}
  V(\phi_0,S) &=& \lambda\left(\phi_0^2-\frac{v_{EW}^2}{2}\right)^2+\frac{a_1}{2}\left(\phi_0^2-\frac{v^2_{EW}}{2}\right)S+\frac{a_2}{2}\left(\phi_0^2+\frac{v^2_{EW}}{2}\right)S^2 \nonumber \\
  && +\frac{1}{4}\left(2b_2+a_2v_{EW}^2\right)S^2+\frac{b_3}{3}S^3+\frac{b_4}{4}S^4,
  \label{singlet}
\end{eqnarray}
where $\phi_0$ is the neutral component of the Higgs doublet and $S$ is the additional real singlet. $\phi_0$ is expressed as $\phi_0=(h+v)/\sqrt{2}$, where $v$ is the vev of the doublet. Similarly, the vev of the singlet is denoted as $x$. In the limit of $(v, x)=(v_{EW},0)$, the EWSB is minimized.

After EWSB, a new Higgs boson, $H_2$, is introduced by diagonalizing the Higgs mass matrix from the gauge eigenstates into the mass eigenstates. The mixing angle $\theta$ and the parameters of Eq. (\ref{singlet}) satisfy the following relations:
\begin{eqnarray}
  a_1 &=& \frac{m_H^2-m^2_{H_2}}{v_{EW}}\sin{2\theta}, \\
  b_2+\frac{a_2}{2}v_{EW}^2 &=& m_H^2\sin^2{\theta}+m_{H_2}^2\cos^2{\theta}, \\
  \lambda &=& \frac{m_H^2\cos^2{\theta}+m_{H_2}^2\sin^2{\theta}}{2v^2_{EW}}.
  \label{singletrelations}
\end{eqnarray}
Above, $m_H=126$ GeV, and $m_{H_2}$ is the mass of $H_2$. Given $(v, x)=(246~\text{GeV},0)$, the remaining free parameters of SM+S are
\begin{center}
  $m_{H_2}$, $\theta$, $a_2$, $b_3$, $b_4$.
\end{center}
After EWSB, the Higgs self-interactions (in the mass eigenstates) of SM+S are given by
\begin{align}
V_\text{self}\supset{}& \frac{\lambda_{111}}{6} H^3 +\frac{\lambda_{211}}{2} H^2 H_2+\frac{\lambda_{221}}{2} H H_2^2+\frac{\lambda_{222}}{6} H_2^3\nonumber\\
+{}&\frac{\lambda_{1111}}{24}H^4+\frac{\lambda_{2111}}{6} H^3 H_2+\frac{\lambda_{2211}}{4} H^2 H_2^2+\frac{\lambda_{2221}}{6} H H_2^3+\frac{\lambda_{2222}}{24}H_2^4.
\end{align}
Expressions for above cubic and quartic couplings in terms of  $m_{H_2}$, $\theta$, $a_2$, $b_3$, and $b_4$ are listed in Ref. \cite{Chen:2014ask}.

\begin{figure}[htbp]
  \centering
      \vspace{0.2cm}
  \subfigure{
  \label{Feynman2.sub.1}
  \includegraphics[width=0.18\textwidth]{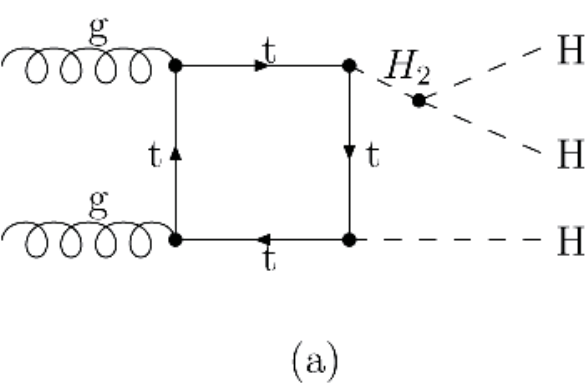}}
  \subfigure{
  \label{Feynman2.sub.2}
  \includegraphics[width=0.18\textwidth]{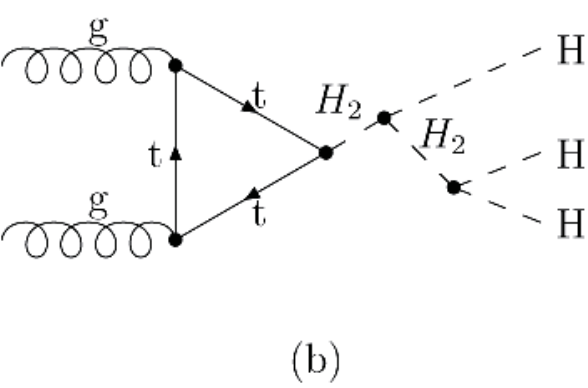}}
  \subfigure{
  \label{Feynman2.sub.3}
  \includegraphics[width=0.18\textwidth]{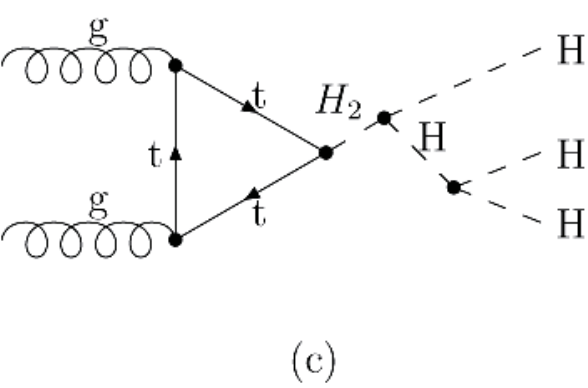}}
  \subfigure{
  \label{Feynman2.sub.4}
  \includegraphics[width=0.18\textwidth]{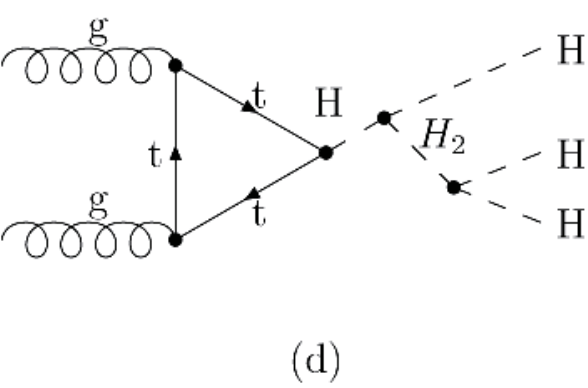}} 
  \subfigure{
  \label{Feynman2.sub.5}
  \includegraphics[width=0.18\textwidth]{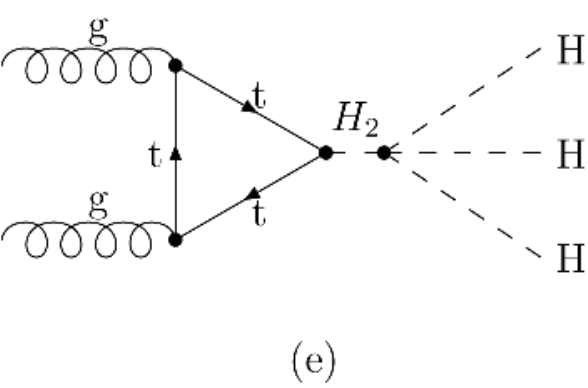}}
        \vspace{-0.2cm}
  \caption{Extra Feynman diagrams which contribute to the process $gg\to HHH$ in the Higgs singlet model are provided.}\label{feynman2}
\end{figure}

The introduction of the heavy Higgs, $H_2$, adds five kinds of diagrams to the process $gg\to HHH$. They are a box quark loop $\to H(H_2)\to H (HH)$ [Fig. \ref{Feynman2.sub.1}]; triangle quark loop $\to H_2\to H  (H_2^*)\to H(HH)$ [Fig. \ref{Feynman2.sub.2}]; triangle quark loop $\to H_2\to H (H^*)\to H(HH)$ [Fig. \ref{Feynman2.sub.3}];  triangle quark loop $\to H \to H (H_2^*)\to H(HH)$ [Fig. \ref{Feynman2.sub.4}]; and the triangle quark loop $\to H_2 \to HHH$ [Fig. \ref{Feynman2.sub.5}]. The first four diagrams all involve the trillinear coupling $\lambda_{211}$. The last diagram instead contains the quartic coupling $\lambda_{2111}$.

We choose benchmark points that introduce a resonance of $H_2\rightarrow HHH$ where the triple-Higgs production is enhanced and other decay channels of $H_2$ are suppressed. Besides, we require the benchmark points satisfy the Higgs vacuum stability requirement; i.e., the Higgs potential at extrema $(v, x)=(v_{EW}, 0)$ is no larger than those at the other eight potential local extrema.\footnote{The nine potential local extrema of the Higgs potential are $(v, x)=(v_{EW}, 0)$, $(-v_{EW}, 0)$, $(v_+, x_+)$, $(-v_+, x_+)$, $(v_-, x_-)$, $(-v_-, x_-)$, $(0, x_1^0)$, $(0, x_2^0)$ and $(0, x_3^0)$. Detailed expressions are given by Eq. (24) and (B1) in Ref. \cite{Chen:2014ask}).}

In the parameter scan, we require 
\begin{equation}
378~ \text{GeV}\leq m_{H_2} \lesssim 2~ \text{TeV},
\label{massres}
\end{equation} 
where the lower limit is set by requiring on-shell triple-Higgs final states and the upper limit is from the perturbative unitarity constraint.  We adopt the restriction $\sin\theta^2 \leq 0.12$ on $\theta$ from fittings of the Higgs coupling strengths~\cite{ATLAS-CONF-2014-010}. We also constrain
\begin{equation}
|a_2|\leq 4\pi,\quad |b_3|/v_{EW}\leq 4\pi, \quad 0<b_4\lesssim 8\pi/3, \quad 0<\lambda\leq 4\pi/3,\quad a_2^2<4 \lambda b_4
\end{equation}
from requirements of perturbative unitarity, perturbativity, and the positivity of the potential. The perturbative unitarity bounds above are obtained as follows. We compute the normalized spherical amplitude matrix for quadratic scattering between $W_L^+W_L^-$, $Z_L Z_L$, $HH$, $HH_2$, and $H_2H_2$. Then, we require the real parts of the eigenvalues of the matrix to be smaller than $1/2$~\cite{Lee:1977eg,Gunion:1989we,Basso:2010jt,Pruna:2013bma}. Under a good approximation, we take the limit $\theta\to 0$. This leads to restrictions $\lambda\lesssim 4\pi/3$ and $b_4 \lesssim 8\pi/3$. The former restriction yields an upper limit on $m_{H_2}$ as shown in Eq. (\ref{massres}).

The benchmark points are listed in Table \ref{benchmarktable} and \ref{benchmarktable2}. 
They are obtained by optimizing the cross section for $pp\to H_2 \to HHH$ under the narrow width approximation [$\sigma(pp \to H_2\to HHH)\approx \sigma(gg\to H_2)\times BR(H_2 \to HHH)$; here, we only consider $H_2$ production via gluon fusion]. We find a maximal triple Higgs production cross section is in coincidence with a minimal $BR(H_2 \to HH)$, 
as demonstrated by bench mark points B1, B2, and B3 in \ref{benchmarktable2}.

\begin{center}
\begin{table}
\begin{center}
\begin{tabular}{|l|c|c|c|}
  \hline
                  & B1      & B2  &  B3        \\
  \hline
  $m_{H_2}$ (GeV) & $460$   & $500$  &  $490$     \\
  $\theta$        & $0.354$  & $0.354$  &  $0.354$    \\
  $a_2$           & $3.29$ & $3.48$  &  $3.43$ \\
  $b_3$  (GeV)  & $-706$   & $-612$  &  $-637$   \\
  $b_4$           & $8.38$  & $8.38$ &  $8.38$   \\
  \hline
\end{tabular}
\end{center}
\caption{The benchmark points to probe the singlet+SM model. \label{benchmarktable}}
\end{table}
\end{center}

\begin{center}
\begin{table}
\begin{center}
\begin{tabular}{|l|c|c|c|}
  \hline
                  & B1      & B2   &  B3                        \\
  \hline
    $\Gamma_\text{tot}(H_2)$ (GeV)  & $5.6$  &  $7.5$  & $7.0$ \\
  \hline
    $BR(H_2\to W^{+} W^{-})$ & $0.57$ & $0.56$& $0.57$\\
    $BR(H_2\to Z Z)$ & $0.27$ & $0.27$        & $0.27$\\
    $BR(H_2\to t \bar t)$ & $0.15$ & $0.16$   & $0.16$\\
    $BR(H_2\to b {\bar b})$ &$3.4 \times 10^{-4}$ & $2.8 \times 10^{-4}$ & $2.9 \times 10^{-4}$\\    
    $BR(H_2\to H H)$ & $5.3 \times 10^{-7}$ & $8.8 \times 10^{-7}$ & $1.5 \times 10^{-7}$ \\
    \hline
    $BR(H_2\to H H H)$  & $1.0 \times 10^{-3}$ & $1.4 \times 10^{-3}$  &  $1.3 \times 10^{-3}$ \\ 
    \hline
    $\sigma(gg \to H_2)$ @ 14 TeV (fb)  &  $3.2\times 10^2$ & $2.3\times 10^2$ & $2.5\times 10^2$\\  
    $\sigma(gg \to H H H)$ @ 14 TeV (fb) & $0.70$ & $0.69$ & $0.71$\\   
    $\sigma(gg \to H_2)$ @ 100 TeV (fb)  &  $1.4\times 10^{4}$ & $1.1\times 10^{4}$ & $1.2\times 10^{4}$\\  
    $\sigma(gg \to H H H)$ @ 100 TeV (fb) & $37$ & $38$ & $39$\\
  \hline
\end{tabular}
\end{center}
\caption{The total width  and branching ratios of $H_2$. The cross sections of $gg \to H_2$ and $gg \to H H H$ are listed to demonstrate the enhancement due to the resonance. \label{benchmarktable2}}
\end{table}
\end{center}

There are a few comments in order on these benchmark points B1, B2 and B3 given in Table \ref{benchmarktable2}:
\begin{enumerate}
\item It is remarkable that the resonance of $H_2$ can enhance the production of triple-Higgs boson final states by $1$ order of magnitude for the benchmark points. 
\item Enhancements in other channels, like $ZZ$, could be marginally feasible at LHC Run 2. Meanwhile, the triple-Higgs boson final states could also be reachable for the LHC high luminosity run. For a 100 TeV collider, both $ZZ$ and triple-Higgs boson final states could be reachable.
\item Enhancements in di-Higgs boson final states can be safely neglected due to the tiny branching fraction of $H_2 \to H H$.
\end{enumerate}

We implement the model based on the loop\_sm module in MadGraph5/aMC@NLO \cite{Alwall:2014hca}.
First, we add the model parameters, then
implement all the relevant vertices and couplings.
As well as the tree-level vertices, the relevant vertices for R2 terms defined in the OPP method \cite{Ossola:2006us} are also added according to Ref. \cite{Draggiotis:2009yb}.

\begin{figure}[htbp]
  \centering
  \subfigure{
  \label{Fig10.sub.1}\thesubfigure
  \includegraphics[width=0.4\textwidth]{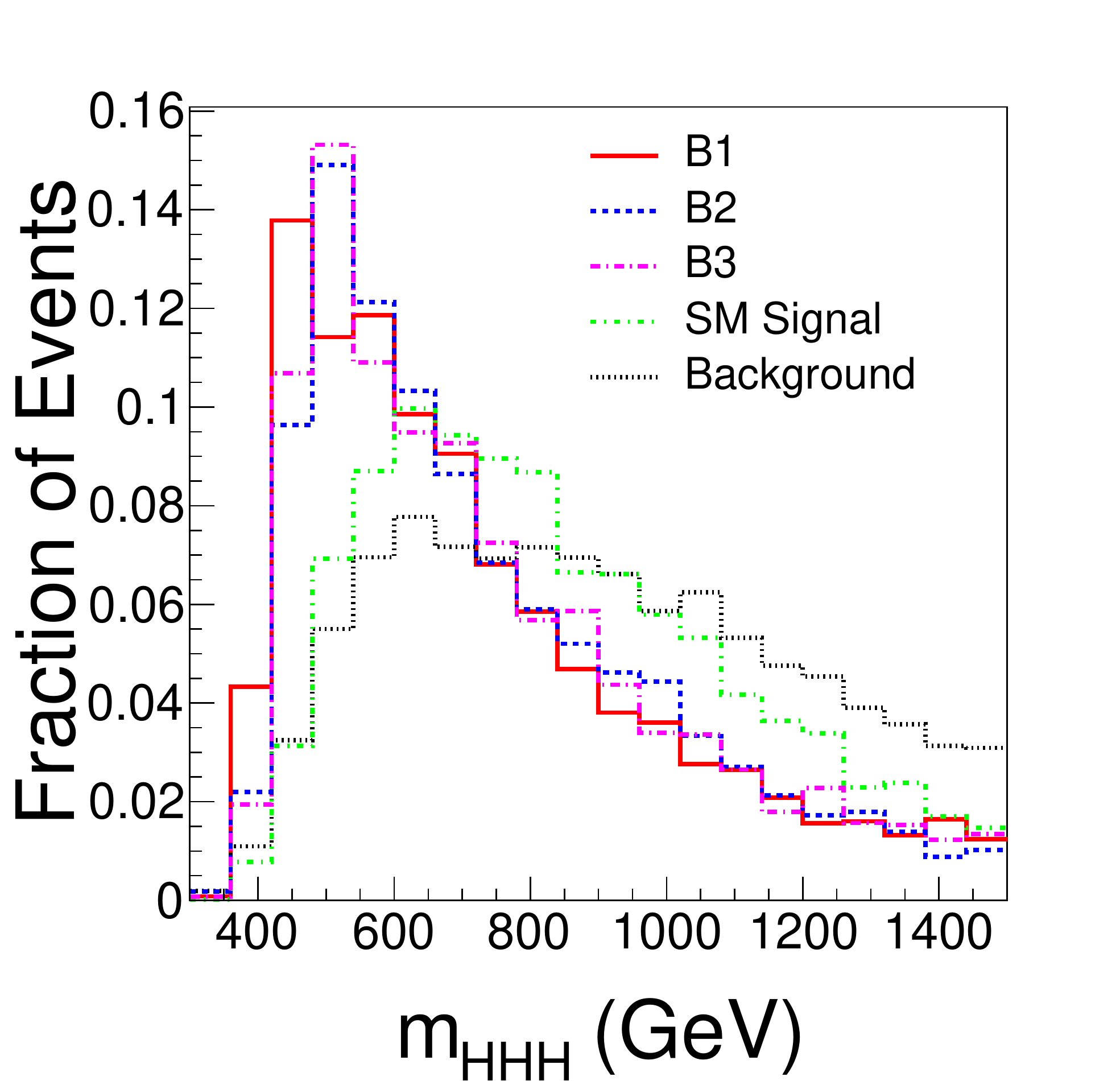}}
  \subfigure{
  \label{Fig10.sub.2}\thesubfigure
  \includegraphics[width=0.4\textwidth]{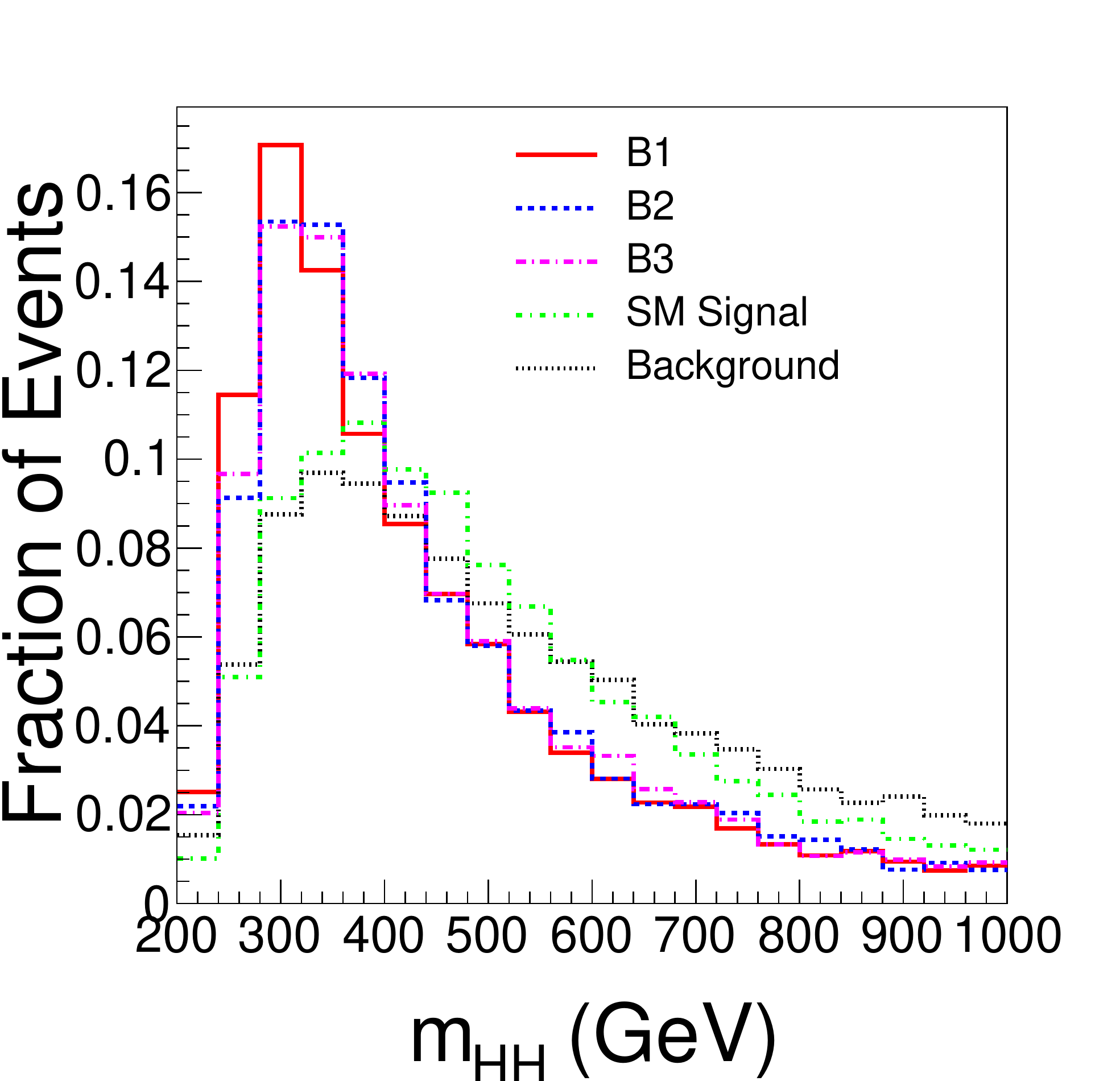}}
  \caption{The detector level distributions of (a) the invariant mass of three Higgs bosons and (b) the invariant mass of di-Higgs bosons on three benchmark points of the singlet+SM model, compared to the distributions of the SM signal and backgrounds.}\label{fig10}
\end{figure}

The triple-Higgs events at this model can be generated efficiently by the new version of MadGraph5/aMC@NLO \cite{Hirschi:2015iia}, which can handle the loop-induced process. To perform the feasibility study, we generate 40,000 events for each benchmark point. We conduct the same analysis as demonstrated in the previous sections. Here, we present our results on these three benchmark points in Fig. \ref{fig10} and Table \ref{singletbdt}. 

Figure \ref{Fig10.sub.1} shows the invariant mass of the triple-Higgs boson on three benchmark points. Comparing to the SM signal and background, the distributions of B1 and B2 have a resonance peaks around $450$ and $500$ GeV, respectively. These peaks are close to the peak from pentegon diagrams, so the resonance peaks are broadened. Fig. \ref{Fig10.sub.2} shows the invariant mass of di-Higgs bosons. When the new diagrams are introduced, the invariant mass of three Higgs bosons tends to be around threshold around $300$ GeV. Because the branching ratio $BR(H_2\to H H)\approx 0$ in B1 and B2, there are not peaks around the mass of $m_{H_2}$.

Table \ref{singletbdt} shows the significances of these three benchmark points. It is observed that the significances can be improved from $0.2$ to $2.1$, $2.5$, and $2.3$, respectively. To obtain these numbers, we estimate the production rate by multiplying the leading-order cross section computed by the MadGraph5 with a $K$ factor extracted from the reference \cite{Dittmaier:2011ti} where N$^3$LO QCD corrections and NLO EW corrections for $gg \to H_2$ have been taken into account. There are two reasons to do so: 1) $H_2$ coupling to the top quark is similar to that of the SM-like Higgs boson, and its coupling strength is equal to $y_t \sin(\theta)/\sqrt{2}$; 2) the contribution of $gg \to H_2 \to HHH$ is the overwhelming process for the triple-Higgs boson production in these benchmark points. As described above, the new resonance can enhance $1$ order of magnitude of the triple-Higgs production rate. Moreover, the new cuts from the invariant mass of triple Higgs and di-Higgs can also improve the discrimination of signal and background events. Therefore, we use the $K$ factor of $gg \to H_2$ to estimate the $K$ factor of $gg \to H_2 \to HHH$. It is noticed that this agrees with the $K$ factor computed in Ref. \cite{Dawson:2015haa}.

\begin{table}
  \begin{center}
  \begin{tabular}{|c|c|c|c|c|}
  \hline
                  &  SM(BDT$>0.02$)                &  B1(BDT$>-0.02$)     &  B2(BDT$>-0.02$)     &  B3(BDT$>-0.03$)\\
  \hline
  Signal          &  $34$                        &  $3.7\times 10^2$             &  $4.4\times 10^2$             &  $4.6\times 10^2$\\
  \hline
  Background      &  $2.8\times{10^4}$             &  $3.0\times{10^4}$   &  $3.1\times{10^4}$   &  $4.0\times{10^4}$\\
  \hline
  $S/B$           &  $1.2\times{10^{-3}}$          &  $1.2\times{10^{-2}}$&  $1.4\times{10^{-2}}$&  $1.1\times{10^{-2}}$ \\
  \hline
  $S/\sqrt{S+B}$  &  $0.20$                        &  $2.1$              &  $2.5$              &  $2.3$\\
  \hline
  \end{tabular}
  \end{center}
  \caption{\label{singletbdt}The numbers of events and the efficiencies of the BDT method on SM and the three benchmark points of the singlet+SM model. Here, the total integrated luminosity is $30$ ab$^{-1}$.}
\end{table}

\section{Discussion and Conclusion}
\label{discussion}

In this paper, we have studied the feasibility of triple-Higgs production via $4b2\gamma$ final states at a 100 TeV hadron collider. We explore some kinematic cuts which can reduce background effectively, and we find it is challenging to measure the quartic coupling of the Higgs boson in the SM even at a 100 TeV hadron collider if luminosity is assumed to be 30 ab$^{-1}$ due to its small cross section and the huge QCD background. To observe the signal of the SM, an integrated luminosity up to $1.8\times 10^4$ ab$^{-1}$ is required. 

If new physics that can enhance the triple-Higgs production rate is taken into account, it is promising to discover triple-Higgs production via the $b\bar{b}b\bar{b}\gamma\gamma$ channel. For the effective Higgs potential model introduced in Eq. (\ref{eq1.1}), we find that $\lambda_3$ can be confined to the range $[-1,5]$ and $\lambda_4$ can be confined to the range $[-20,30]$. 

In our detector simulation, we have assumed that $b$-tagging efficiency is at most around $60\%$. According to the current results from both CMS and ATLAS collaborations, the $b$-tagging efficiency can reach up to around $70\%$. Therefore, we can expect that a better result could be yielded when a larger $b$-tagging efficiency is taken.

In the analysis presented in Sections \uppercase\expandafter{\romannumeral3}--\uppercase\expandafter{\romannumeral5}, we have applied a $b$-tagging cut at $n_b\geq 2$. We also expose other $n_b$ cases in Table \ref{btag2}. It is found that the analysis with either $n_b\geq 2$ or $n_b\geq 3$ is the best. For $n_b\geq 3$, the signal events are lost by a factor of 60\%, but the background events $pp\to b\bar{b}jj\gamma\gamma$ and $pp\to H(\gamma\gamma)t\bar{t}$ are suppressed by $1$ order of magnitude. Although the background $pp\to b\bar{b}b\bar{b}\gamma\gamma$ becomes as important as $pp\to H(\gamma\gamma)t\bar{t}$, we obtain a better $S/B$ and $S/\sqrt{S+B}$.

Although most of the signal events are kept for $n_b\geq 1$,   backgrounds there are substantial. They are three times larger than those for $n_b\geq 2$. Besides, QCD contributes a huge background of  $4j 2\gamma$ with one light jet faking a $b$ jet. On the other extreme, $n_b\geq 4$ can effectively suppress the background [a factor of $\mathcal O (10)$ less than $n_b\geq 3$]. But the signal then suffers a huge loss that leads to a low significance. Analysis of the case $n\geq 4$ should only be considered if the production rate of the signal is sufficiently large, such as in the singlet+SM model.

\begin{figure}[htbp]
  \centering
  \subfigure{
  \label{Fig11.sub.1}\thesubfigure
  \includegraphics[width=0.45\textwidth]{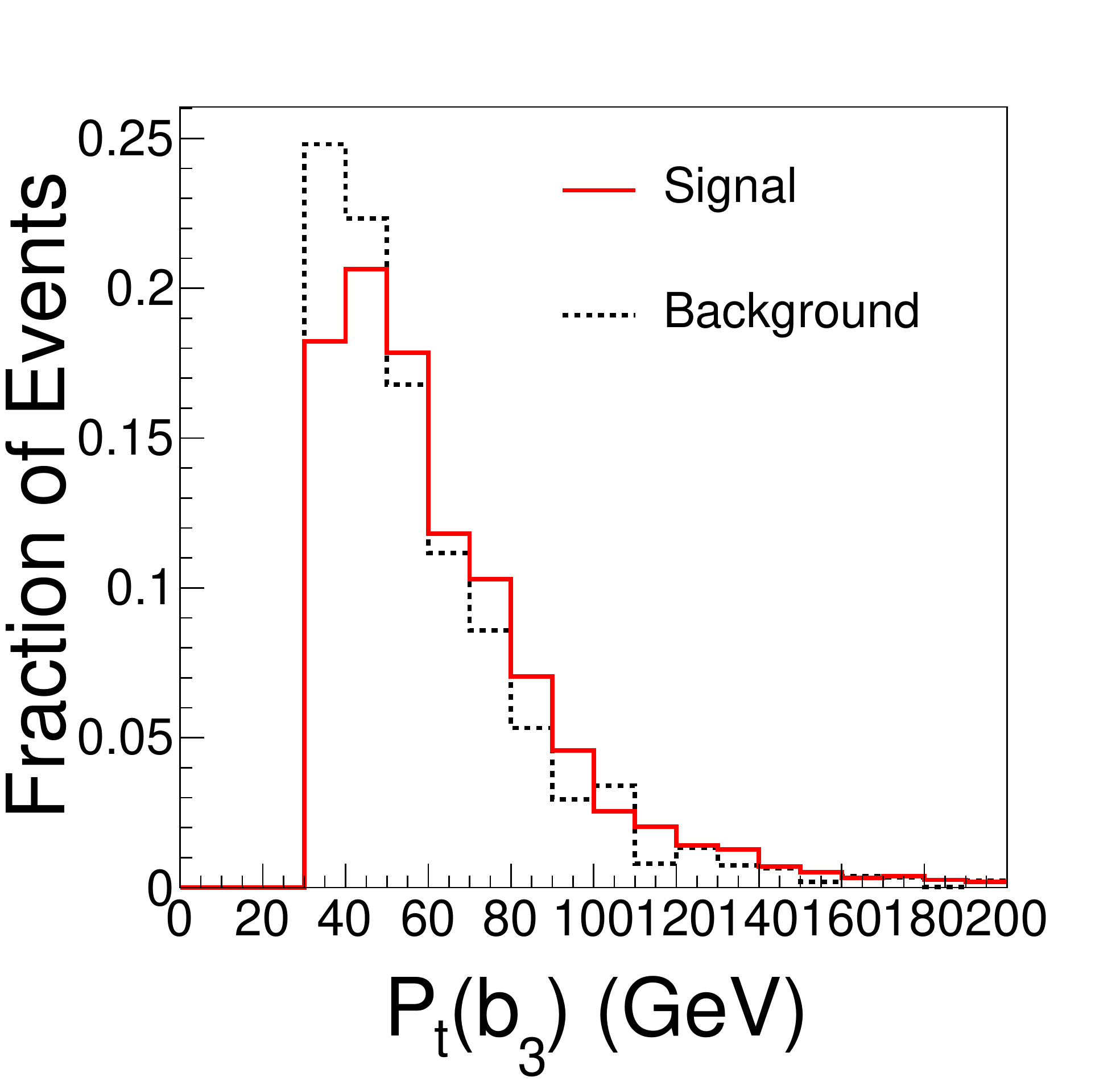}}
  \subfigure{
  \label{Fig11.sub.2}\thesubfigure
  \includegraphics[width=0.45\textwidth]{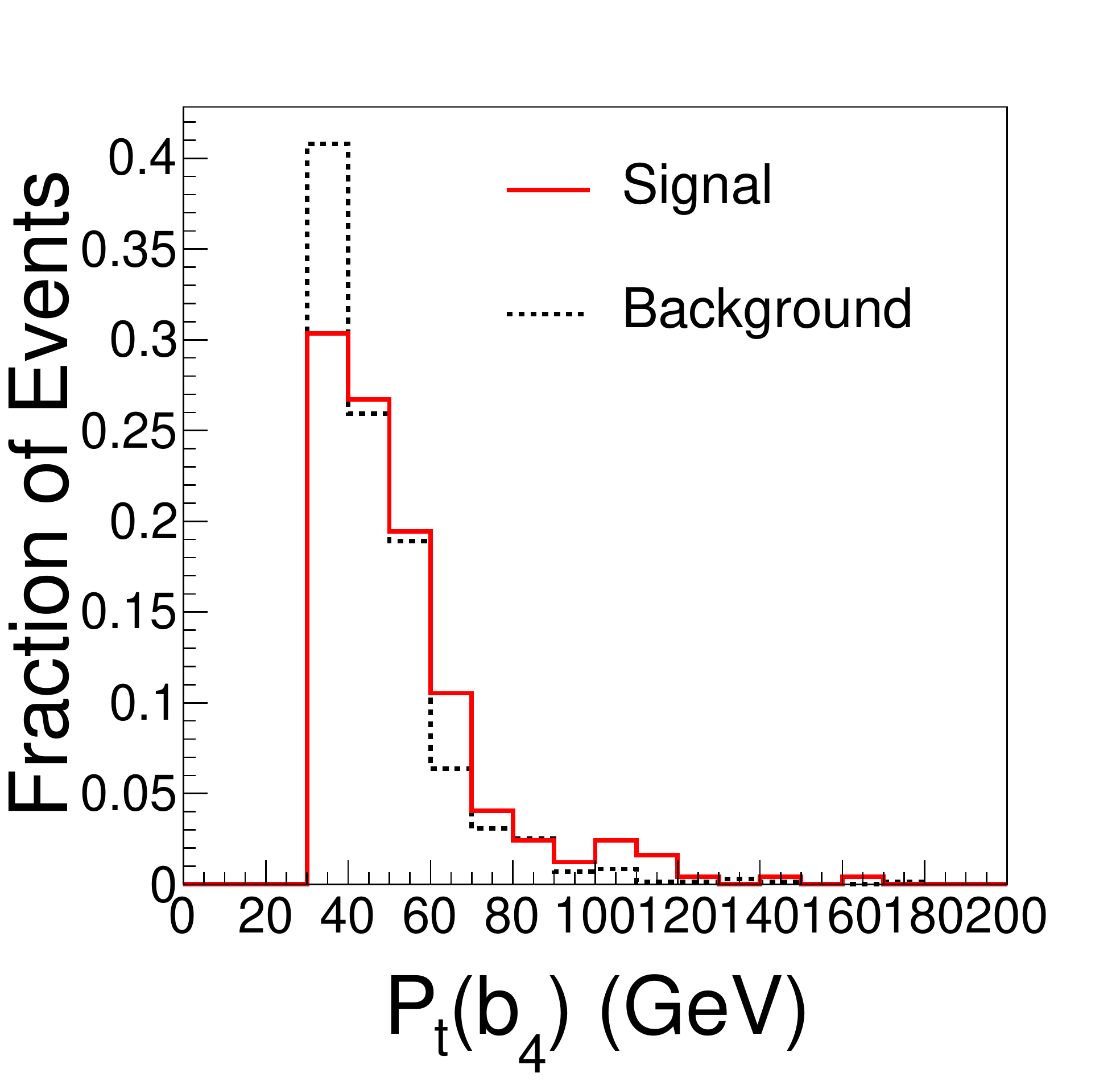}}
  \caption{The transverse momentum distributions of (a)  the third and (b) the fourth hardest tagged $b$ jets. }\label{fig11}
\end{figure}

It is interesting to explore the underlying reasons for the loss of signal events in both $n_b\geq 3$ and $n_b \geq 4$ analyses. Such a loss can be expected from the $b$-tagging efficiency characterized by Eq. (\ref{eq6}). One finds that the hardest $b$-tagging jet has a peak around $120$ GeV, while the second hardest jet has a peak around $50$ GeV. Based on Eq. (\ref{eq6}), the $b$-tagging efficiency $\epsilon_b$ reduces to $0.4$ when $P_t(j)\sim 50$ GeV. In the events with three or more $b$-tagged jets, the third hardest jet has a transverse momentum less than $50$ GeV, and $\epsilon_b$ is further reduced, which leads to a $50\%$ loss of signal events. It becomes even worse when we require $n_b\geq 4$, where the peak of the transverse momentum of the fourth hardest jet is less than 30 GeV and the $b$-tagging efficiency is dropped down to less than 0.3, as demonstrated in Table \ref{tab:btag}. Fig. \ref{fig11} shows the transverse momentum of the third and fourth hardest tagged $b$ jets, which provide evidence why the signal events suffer a big loss when we increase the number of tagged $b$ jets. It will be greatly helpful for the triple-Higgs discovery if the detectors of future colliders can improve the $b$-tagging efficiency for soft $b$ jets.
\begin{center}
\begin{table}
\begin{center}
\begin{tabular}{|c|c|c|c|c|}
  \hline
                           & $n_b\geq 1$        & $n_b\geq 2$        & $n_b\geq 3$      &$n_b\geq 4$    \\
  \hline
  SM signal                & $79$             & $50$             & $18$           &$2.8$\\
  $b\bar{b}jj\gamma\gamma$       & $7.0\times{10^5}$  & $2.3\times{10^5}$  & $1.8\times{10^4}$&$850$\\
  $H(\gamma\gamma)t\bar{t}$& $7.0\times{10^4}$  & $2.2\times{10^4}$  & $1.7\times{10^3}$&$21$       \\
  $b\bar{b}b\bar{b}\gamma\gamma$       & $5.1\times{10^3}$  & $3.6\times{10^3}$  & $1.4\times{10^3}$&$260$       \\
  \hline
  $S/B$                    & $1.0\times{10^{-4}}$ & $1.9\times{10^{-4}}$&$8.5\times{10^{-4}}$&$2.5\times{10^{-3}}$\\
  $S/\sqrt{S+B}$           & $8.9\times{10^{-2}}$ & $9.8\times{10^{-2}}$&$0.12$&$8.3\times{10^{-2}}$\\
  \hline
\end{tabular}
\end{center}
\caption{\label{btag2}The significances for analyses with different numbers of tagged $b$ jets. Here, the luminosity is $30$ ab$^{-1}$.}
\end{table}
\end{center}

We find Ref. \cite{Papaefstathiou:2015paa} has done a similar study on triple-Higgs productions but with only the case $n_b\geq 4$ considered. The authors show that a signal-to-background ratio can reach $\sim 1$ at a 100 TeV hadron collider, which requires a high $b$-tagging efficiency ($80\%$), a low light-jet mistagging rate ($1\%$), and excellent photon identification. 
We have focused on the case $n_b \geq 2$ instead. We show that an important background $pp\to bbjj\gamma\gamma$ could contribute significantly in those cases $n_b \geq 2$, $n_b \geq 3$, and $n_b \geq 4$. Our results indicate that this type of background events are important in the analysis of  $p p \to HHH\to b\bar{b}b\bar{b}\gamma\gamma$ channel and could contribute to $\sim 33\%$ of the total background events. Meanwhile, our results also show that the $b$-tagging to soft jets is crucial to discover the signal.
Meanwhile, the process $ p p \to  t {\bar t} H$ can contribute around $30\%$ of the total background of the SM in the case $n_b\geq 2$. After taking into account more realistic $b$-tagging efficiency, especially those soft $b$ jets in signal events, our analysis shows that the discovery of the signature of the triple-Higgs final state in the SM is indeed challenging. 

In the model where an extra Higgs singlet is added to the SM, we propose a few benchmark points where the production rate of $gg\to HHH$ can be enhanced dramatically by new resonances. Because of the existence of resonances, we can have more efficient kinematic cuts to suppress the SM background. In our work, the efficiency can be up to $2.5$ on benchmark point B2 when the luminosity is $30$ ab$^{-1}$. 

In our analysis, the $K$ factor of $2b2j2\gamma$ is assumed to be 1. We may also use the result computed for the process $pp \to 4b$ \cite{Worek:2013zwa} to estimate it, where the $K$ factor is around 1.4. Since this is the main background for the signal channel, our results could be significantly affected by this factor. But our results could serve as a guide to estimate the required luminosity. Meanwhile, this work indicates that the QCD corrections of the process $pp \to 2b2j2\gamma$ could be important for triple-Higgs production and should be studied carefully.

Here, we would like to address the fake photon issue. The high-energy neutral pions can fake photons in the electromagnetic calorimeter (ECAL). The cross sections of the processes $pp \to 2b 4j$ and $pp\to2b3j\gamma$ are found by using Alpgen\cite{Alpgen} to be 2.1 $\times$ 10$^{5}$ and 250 pb, respectively. When the fake photon rate is assumed to be $0.1\%$, the cross sections are dropped down to 1260 and 750 fb (combinatorial factors have been taken into account), respectively. After the invariant mass window cut on the diphoton invariant mass, we noticed that only around $10\%$ events can contribute like events $2b2j 2 \gamma$. Then we noticed that the combination of these types of background can be of the same size as $pp\to 2b2j 2 \gamma$. This will make the minimum luminosity even larger by the number estimated in Table \ref{tab:table4}. The minimum luminosity derived from the results of mode $n_b \geq 3$ could be more robust  than that of the mode $n_b \geq 2$ after taking into account the contribution of fake photon events to the main background $2b2j2\gamma$ and the minimal luminosity close to that quoted in Table \ref{tab:table4}. If the fake rate can be further reduced experimentally, then combining both $n_b =  2$ and $n_b = 3$ modes gains us a little in reducing the minimal required luminosity.

The next step of this work is to study the feasibility of other channels, either in the SM or new physics models. The potential discovery channels and their branching ratios for triple-Higgs production are listed in Table \ref{tab:table5}. One can find that the $b\bar{b}b\bar{b}W^+W^-$ channel has the largest branching ratio and the number of signal events should be increased dramatically. However, the SM backgrounds might be too large for this channel. For example, the cross section of $pp \rightarrow b\bar{b}t\bar{t}$ can be up to $\sim 10^3$ pb, and it could be difficult to reduce such a large background. For the same reason, the $HHH\rightarrow b\bar{b}b\bar{b}b\bar{b}$ channel might also be difficult, unless we can find a better way to suppress the background. The channels with more than 4 $W$ bosons might also be feasible. For highly boosted Higgs bosons in the triple-Higgs boson final states, the jet substructure techniques, like Higgs-tagger methods \cite{Butterworth:2008iy}, could also be investigated. These studies will be carried out in our future projects.

\begin{center}
\begin{table}
  \begin{center}
  \begin{tabular}{|l|c|}
  \hline
  Decay channel                                         &  Branching ratio      \\
  \hline
  $HHH\rightarrow b\bar{b}b\bar{b}W^+W^-$               &  $22.34\%$          \\
  \hline
  $HHH\rightarrow b\bar{b}b\bar{b}b\bar{b}$             &  $20.30\%$          \\
  \hline
  $HHH\rightarrow b\bar{b}W^+W^-W^+W^-$                 &  $8.20\%$          \\
  \hline
  $HHH\rightarrow b\bar{b}b\bar{b}\tau^+\tau^-$         &  $7.16\%$           \\
  \hline
  $HHH\rightarrow b\bar{b}b\bar{b}gg$                   &  $6.54\%$           \\
  \hline
  $HHH\rightarrow b\bar{b}b\bar{b}ZZ$                   &  $2.69\%$           \\
  \hline
  $HHH\rightarrow W^+W^-W^+W^-W^+W^-$                   &  $1.00\%$          \\
  \hline
  $HHH\rightarrow W^+W^-W^+W^-\tau^+\tau^-$             &  $0.96\%$           \\
  \hline
  $HHH\rightarrow W^+W^-W^+W^-gg$                       &  $0.88\%$           \\
  \hline
  $HHH\rightarrow W^+W^-W^+W^-ZZ$                       &  $0.36\%$           \\
  \hline
  $HHH\rightarrow b\bar{b}b\bar{b}\gamma\gamma$         &  $0.29\%$           \\
  \hline
  \end{tabular}
  \end{center}
  \caption{\label{tab:table5} Some possible discovery channels for triple-Higgs production are listed. Channels with branching fraction less than $0.1\%$ are omitted.}
\end{table}
\end{center}

\begin{acknowledgments} 
We thank Sally Dawson and Giovanni Marco Pruna for helpful discussions and Andreas Papaefstathiou for useful communication. The work of Qi-Shu Yan and Xiaoran Zhao are supported by the Natural Science Foundation of China under the Grants No. 11175251 and No. 11475180. The work of Chien-Yi Chen is supported in part by the U.S. Department of Energy under Grant No. DE-AC02-98CH10886 and Contract No. DE-AC02-76SF00515 and by NSERC, Canada. Research at the Perimeter Institute is supported in part by the Government of Canada through NSERC and by the Province of Ontario through MEDT. Yi-Ming Zhong is supported by the U.S. Department of Energy Grant No. DESC0008061.
\end{acknowledgments}

\appendix 
\section{Setup for the detector simulation}
\label{detectorsimulation}
 In the detector simulation, the radius and half-length of the magnetic field coverage are assumed to be $3.0$ and $5.0$ m, respectively. The axial magnetic field is $5.0$ T. The energy resolution formula of an ECAL is assumed to be 
\begin{equation}
   \sigma_{ECAL}=
   \begin{cases}
   \sqrt{0.007^2\left(\frac{E}{\text{GeV}}\right)^2 + 0.07^2\left(\frac{E}{\text{GeV}}\right) + 0.35^2}, &\mbox{if $|\eta|\leq 3.0$} \, ,\\
   \sqrt{0.107^2 \left(\frac{E}{\text{GeV}}\right)^2 + 2.08^2 \left(\frac{E}{\text{GeV}}\right)}. &\mbox{if $3.0<|\eta|\leq 5.0$}\,,\\
   \end{cases}
\end{equation}
The energy resolution formula for a hadron calorimeter (HCAL) is assumed to be
\begin{equation}
   \sigma_{HCAL}=
   \begin{cases}
   \sqrt{0.05^2 \left(\frac{E}{\text{GeV}}\right)^2 + 1.5^2 \left(\frac{E}{\text{GeV}}\right)}, &\mbox{if $|\eta|\leq 3.0$}\,,\\
   \sqrt{0.13^2 \left(\frac{E}{\text{GeV}}\right)^2 + 2.7^2 \left(\frac{E}{\text{GeV}}\right)}, &\mbox{if $3.0<|\eta|\leq 5.0$}\,,\\
     0, & \mbox{otherwise.}
   \end{cases}
\end{equation}
Here, $\sigma_{ECAL}$ and $\sigma_{HCAL}$ are the resolutions of ECAL and HCAL, respectively. They are functions of energy, $E$, and pseudorapidity, $\eta$, of charged leptons and jets, respectively. In these formulas, the coefficients are taken from the default CMS card in DELPHES, but the regions of $\eta$ for leptons and jets are extended from $\pm 2.5$ to $\pm 5.0$.

 Details for the lepton detection are listed as follows. The electron efficiency is $95\%$ when $P_t(e)>10$ GeV and $|\eta(e)|\leq 2.5$ but decreases to $85\%$ when $2.5<|\eta(e)|\leq 5.0$. For muons, the efficiency is $95\%$ when $10$ GeV$<P_t(\mu)\leq 1$ TeV and $|\eta(\mu)|\leq 5.0$. When $P_t(\mu)>1$ TeV, the muon efficiency satisfies $0.95\exp[0.5 - P_t(\mu)\times 5.0\times{10^{-4}}]$. The photon efficiency is found to be close to the electron efficiency.

\section{$b$-tagging efficiency curves}
\label{btag}
We adopt the $b$-tagging efficiency curve at the 60\% $b$-jet efficiency working point. It is given by
\begin{equation}
  \epsilon_{b}=
   \begin{cases}\label{eq6}
  0.6\tanh\left[0.03 \left(\frac{P_t(j)}{\text{GeV}}\right)-0.4\right], &\mbox{for $|\eta(j)|\leq 2.5$},\\
  0.5\tanh\left[0.03 \left(\frac{P_t(j)}{\text{GeV}}\right)-0.4\right], &\mbox{for $2.5<|\eta(j)|\leq 5.0$},\\
  0, & \text{otherwise.}
   \end{cases}
\end{equation}
The corresponding mistagging rate of the charm quark is 
\begin{equation}
\epsilon_{c\to b} = 
 \begin{cases}
 0.1\tanh\left[0.03\left(\frac{P_t(j)}{\text{GeV}}\right)-0.4\right],& \text{for $|\eta(j)|\leq 5.0$},\\
   0,& \text{otherwise.}
    \end{cases}
\end{equation}
And the corresponding mistagging rate of light quarks and gluons is
\begin{equation}
\epsilon_{j\to b} = 
 \begin{cases}
0.001, &\text{for $|\eta(j)|\leq 5.0$},\\
   0,& \text{otherwise.}
       \end{cases}
\end{equation}
The light quarks have a small mistagging rate $\epsilon_{j\to b} = 0.001$ for $|\eta(j)|\leq 5.0$.

In Table \ref{tab:btag}, we show how $b$-tagging efficiency varies with reference to the transverse momentum and $\eta$ of jets. We would like emphasize that when the transverse momentum of a $b$ jet is soft, the tagging efficiency is low.
\begin{center}
\begin{table}
  \begin{center}
  \begin{tabular}{|c|c|c|}
  \hline
$P_t$ (GeV) & $\epsilon_b (|\eta(j)|\leq 2.5)$  & $\epsilon_b (2.5 \leq |\eta(j)|\leq 5)$  \\ \hline
120 &  0.60  & 0.50 \\ 
100 & 0.59 & 0.49 \\
80 & 0.58 & 0.48 \\
50 & 0.48 & 0.40 \\
30 & 0.28 & 0.23 \\
\hline
  \end{tabular}
  \end{center}
  \caption{\label{tab:btag} The $b$-tagging efficiency varying with $P_t$ is presented.}
\end{table}
\end{center}

\end{document}